\RequirePackage[mathlines]{lineno} 
\pdfoutput=1 
\documentclass[cits]{JINST}
\usepackage{graphicx}	
\usepackage{amsmath}
\usepackage{amssymb}
\usepackage{subfigure}

\newcommand{\up}[1]{$^{#1}$}

\newcommand{\powero}[1]{\mbox{10$^{#1}$}}
\newcommand{\powert}[2]{\mbox{#2$\times$10$^{#1}$}}

\newcommand{\um}{$\mu$\mbox{m}}

\newcommand{\gev}{\mbox{GeV/}$c^2$}

\newcommand{\ironfive}{\mbox{$^{55}$Fe}}
\newcommand{\ame}{\mbox{$^{241}$Am}}
\newcommand{\poten}{$^{210}$Po}
\newcommand{\poeight}{$^{218}$Po}
\newcommand{\pbten}{$^{210}$Pb}
\newcommand{\biten}{$^{210}$Bi}
\newcommand{\ptwo}{\mbox{$^{32}$P}}
\newcommand{\sitwo}{\mbox{$^{32}$Si}}
\newcommand{\indium}{\mbox{$^{115}$In}}
\newcommand{\urafour}{\mbox{$^{234}$U}}
\newcommand{\radium}{\mbox{$^{226}$Ra}}
\newcommand{\ura}{\mbox{$^{238}$U}}
\newcommand{\tho}{\mbox{$^{232}$Th}}
\newcommand{\radon}{\mbox{$^{222}$Rn}}
\newcommand{\thoron}{\mbox{$^{220}$Ra}}

\title{Measurement of radioactive contamination \\in the high-resistivity silicon CCDs of the DAMIC experiment}

\author{
A.~Aguilar-Arevalo$^a$, 
D.~Amidei$^b$,
X.~Bertou$^c$,
D.~Bole$^b$,
M.~Butner$^{d, j}$,
G.~Cancelo$^d$,
A.~Casta\~{n}eda~V\'{a}zquez$^a$,
A.E.~Chavarria$^e$\thanks{Corresponding author.},
J.R.T.~de~Mello~Neto$^f$,
S.~Dixon$^e$,
J.C.~D'Olivo$^a$,
J.~Estrada$^d$, 
G.~Fernandez~Moroni$^d$,
K.P.~Hern\'{a}ndez~Torres$^a$,
F.~Izraelevitch$^d$,
A.~Kavner$^b$, 
B.~Kilminster$^g$,
I.~Lawson$^h$,
J.~Liao$^g$,
M.~L\'opez$^i$,
J.~Molina$^i$, 
G.~Moreno-Granados$^a$,
J.~Pena$^e$,
P.~Privitera$^e$,
Y.~Sarkis$^a$, 
V.~Scarpine$^d$,
T.~Schwarz$^b$,
M.~Sofo~Haro$^c$, 
J.~Tiffenberg$^d$,
D.~Torres~Machado$^f$,
F.~Trillaud$^a$,
X.~You$^f$ and
J.~Zhou$^e$\\
\llap{$^a$} Universidad Nacional Aut{\'o}noma de M{\'e}xico, M{\'e}xico D.F., M{\'e}xico \\  
\llap{$^b$} University of Michigan, Department of Physics, Ann Arbor, MI, United States \\  
\llap{$^c$} Centro At\'omico Bariloche - Instituto Balseiro, CNEA/CONICET, Argentina \\ 
\llap{$^d$} Fermi National Accelerator Laboratory, Batavia, IL, United States \\
\llap{$^e$} Kavli Institute for Cosmological Physics and The Enrico Fermi Institute, The University of Chicago, Chicago, IL, United States \\
\llap{$^f$} Universidade Federal do Rio de Janeiro, Instituto de  F\'{\i}sica, Rio de Janeiro, RJ, Brazil \\ 
\llap{$^g$} Universit{\"a}t Z{\"u}rich Physik Institut, Zurich, Switzerland \\   
\llap{$^h$} SNOLAB, Lively, ON, Canada \\  
\llap{$^i$}Facultad de Ingenier\'{\i}a - Universidad Nacional de Asunci\'on, Paraguay \\ 
\llap{$^j$} Northern Illinois University, DeKalb, IL, United States\\
E-mail: \email{alvaro@kicp.uchicago.edu}}

\abstract{We present measurements of radioactive contamination in the high-resistivity silicon charge-coupled devices (CCDs) used by 
the DAMIC experiment to search for dark matter particles. Novel analysis methods, which exploit the unique spatial resolution of CCDs, were developed to identify $\alpha$ and $\beta$ particles. Uranium and thorium contamination in the CCD bulk was measured through $\alpha$ spectroscopy, with an upper limit on the \ura\ (\tho) decay rate of 5 (15)\,kg$^{-1}$\,d$^{-1}$ at 95\% CL. We also searched for pairs of spatially correlated electron tracks separated in time by up to tens of days, as expected from \sitwo\,--\ptwo\  or \pbten\,--\biten\ sequences of $\beta$ decays.  The decay rate of \sitwo\ was found to be $80^{+110}_{-65}$\,kg$^{-1}$\,d$^{-1}$ (95\% CI). An upper limit of $\sim$35\,kg$^{-1}$\,d$^{-1}$ (95\% CL) on the \pbten\ decay rate was obtained independently by $\alpha$ spectroscopy and the $\beta$ decay sequence search. These levels of radioactive contamination are sufficiently low for the successful operation of CCDs in the forthcoming 100\,g DAMIC detector.}


\begin{document}

\section{Introduction}

The 
DAMIC (Dark Matter in CCDs) experiment~\cite{Barreto2012264, Chavarria201521} employs the bulk silicon of scientific-grade charge-coupled devices (CCDs) to detect coherent elastic scattering of Weakly-Interacting Massive Particles (WIMPs) \textemdash\ putative yet-to-be-discovered particles which may explain the dark matter in the universe~\cite{Kolb:1990vq, Griest:2000kj, Zurek:2013wia}. By virtue of the low readout noise of the CCDs and the relatively low mass of the silicon nucleus, DAMIC is particularly sensitive to low mass ($<$20\,\gev) WIMPs, which induce nuclear recoils of keV-scale energies. 

As for any direct dark matter search, the ultimate sensitivity of the experiment is determined by the rate of radioactive background that mimics the nuclear recoil signal from WIMPs. At the lowest energies ($<$1\,keV), CCDs, as well as any other present detector technology,  lack the capability of effective discrimination between signals produced by nuclear recoils and those from ionizing electrons, making energy deposits in the active target from $\beta$s and $\gamma$-rays potential backgrounds for a WIMP search.

To suppress such potential backgrounds a breadth of strategies have been adopted. Direct dark matter search experiments are deployed in deep underground laboratories, to eliminate cosmogenic backgrounds. DAMIC is located in the SNOLAB laboratory, 2 km below the surface in the Vale Creighton Mine near Sudbury, Ontario, Canada. Dedicated screening and selection of detector shielding materials, as well as radon-suppression methods, are extensively employed to decrease the background from radioactive decays in the surrounding environment. However, it is the radioactive contamination of the active target that will often determine the feasibility and scalability of different technologies for rare-event searches. Thus, the measurement of the intrinsic contamination of the detector is fundamental, especially for solid-state devices, whose active target cannot be purified after fabrication. Particularly relevant for DAMIC and other silicon-based experiments (e.g. SuperCDMS~\cite{Rau:2012eg}) is the cosmogenic isotope \sitwo, which could be present in the bulk of the detector. Its $\beta$ decay spectrum extends to the lowest energies and may ultimately become an irreducible background.

The dark matter search will be performed with DAMIC100, a detector with 100\,g of sensitive mass. We aim for background rates in the WIMP search region of $\sim$2\,keV$^{-1}$\,kg$^{-1}$\,d$^{-1}$, comparable to those of state-of-the-art ionization detectors~\cite{Aalseth:2012if, Agnese:2013jaa}. This will allow us to probe, after one year of operation, WIMP-nucleon spin-independent interaction cross-sections as small as $10^{-5}$\,pb for WIMPs with masses as low as 2\,GeV/$c^2$. To achieve this goal, the total decay rate of $\beta$ emitters in the bulk of the CCD must be $\sim$300\,kg$^{-1}$\,d$^{-1}$ or lower.

In this paper, we present novel techniques for the measurement of radioactive contamination in the bulk silicon and on the surface of DAMIC CCDs. We exploit the superb spatial resolution of the CCD to derive distinctive signatures for $\alpha$ and $\beta$ particles, and to identify radioactive decay sequences with time separation of up to weeks. 

We describe the properties of the CCDs and the DAMIC setup at SNOLAB in Section~\ref{sec:damic}. Characteristic features of $\alpha$-induced signals, and limits on uranium and thorium contamination through $\alpha$ spectroscopy are presented in Section~\ref{sec:alpha}. We detail in Section~\ref{sec:spatcoinc} a search for spatially correlated $\beta$ decay sequences and corresponding results on \sitwo\ and \pbten\ contamination. Conclusions are drawn in Section~\ref{conclusions}.

\section{The DAMIC detector}
\label{sec:damic}

\subsection{Characteristics of DAMIC CCDs}
\label{sec:overview}

\begin{figure}[t!]
\begin{center}
\subfigure[A CCD pixel]{
\includegraphics[width=0.32\textwidth]{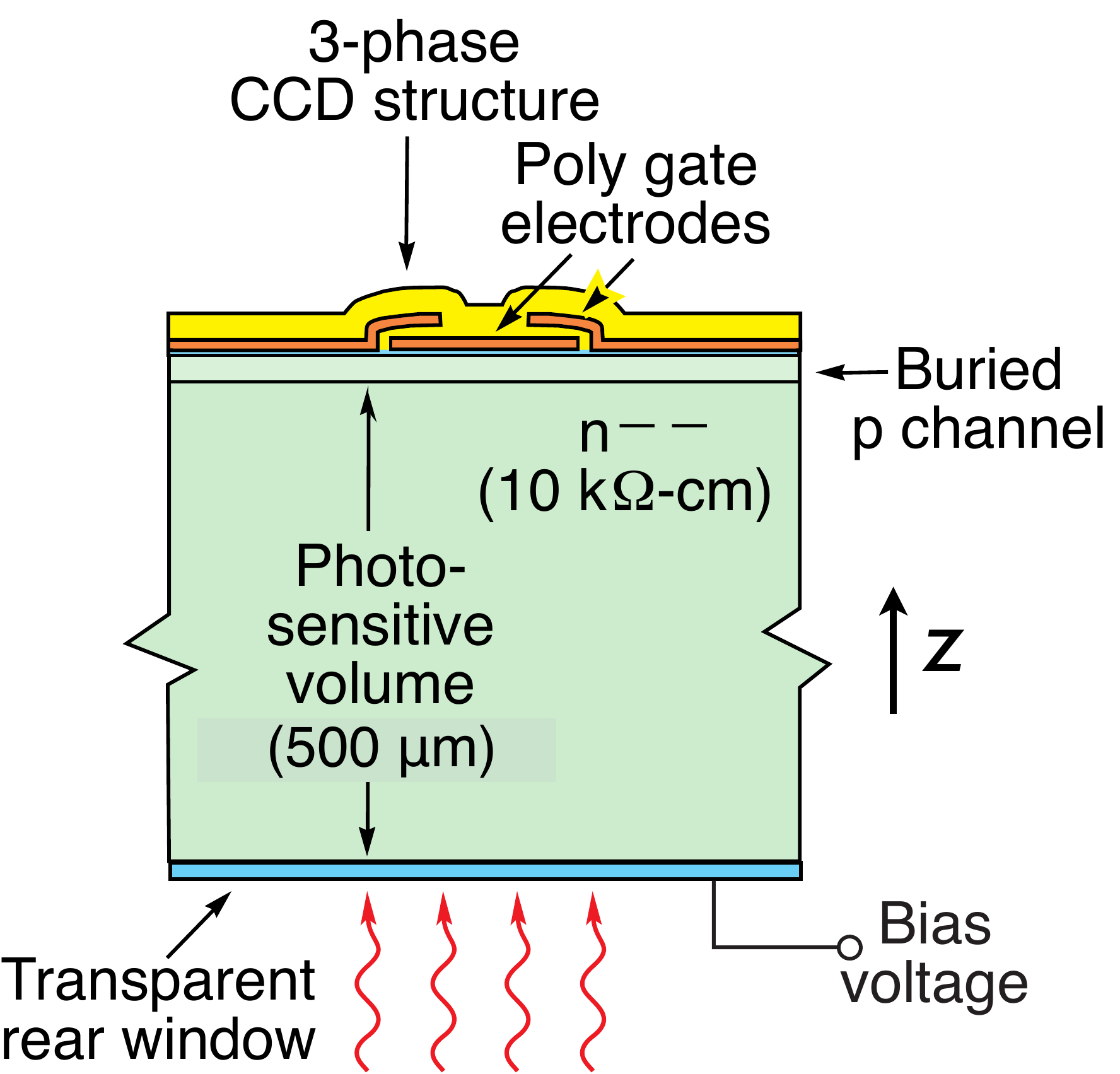}
\label{fig:pixel_sketch}
}
\hspace{4mm}
\subfigure[WIMP detection in a CCD]{
\includegraphics[width=0.59\textwidth]{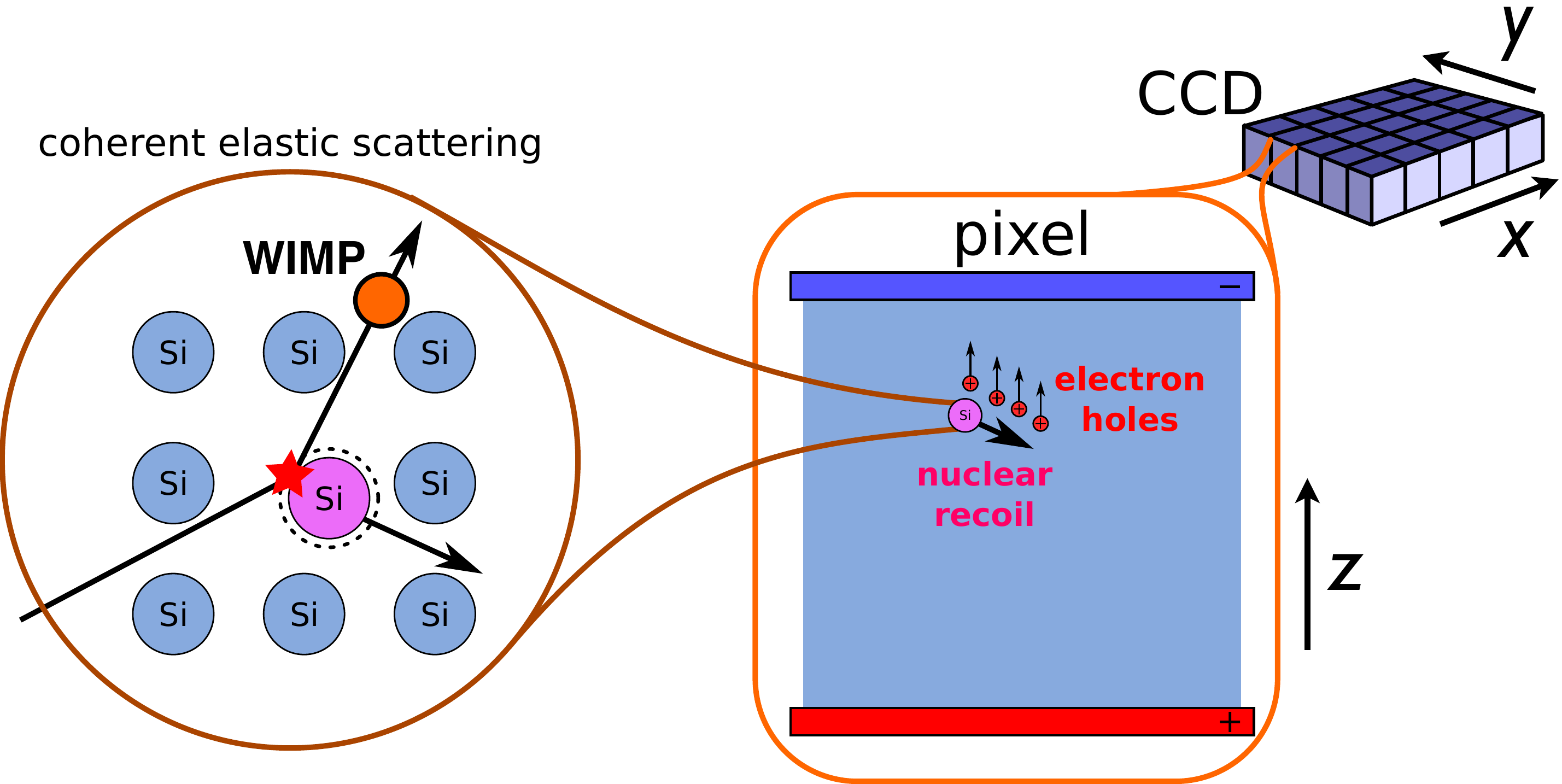}
\label{fig:ccd_sketch}
}
\end{center}
\caption{a)~Cross-sectional diagram of a 15\,$\mu$m\,$\times$\,15\,\um\ pixel in a fully depleted, back-illuminated CCD. The thickness of the gate structure and the backside ohmic contact are $\leq$2\,\um. The transparent rear window, essential for astronomy applications, has been eliminated in the DAMIC CCDs. b)~Dark matter detection in a CCD. A WIMP scatters with a silicon nucleus producing ionization in the CCD bulk. The charge carriers are then drifted along the $z$-direction and collected at the CCD gates.
\label{fig:pixel_ccd_sketch}}

\vspace{0.3 cm}
\begin{center}
\subfigure[Portion of a DAMIC image]{
\includegraphics[width=0.4\textwidth]{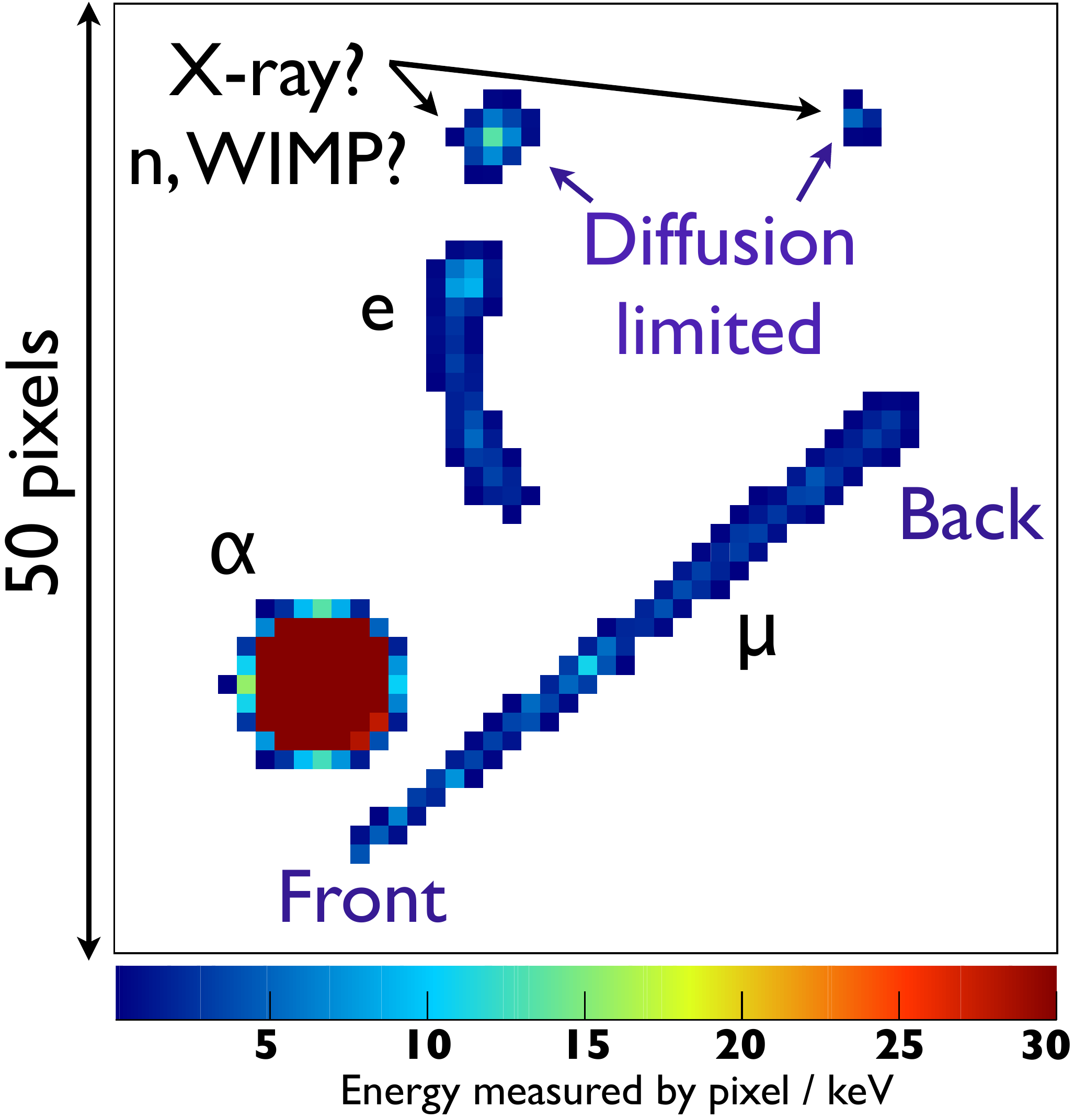}
\label{fig:image_eg}
}
\hspace{2mm}
\subfigure[Emission of Si fluorescence X-ray]{
\includegraphics[width=0.35\textwidth]{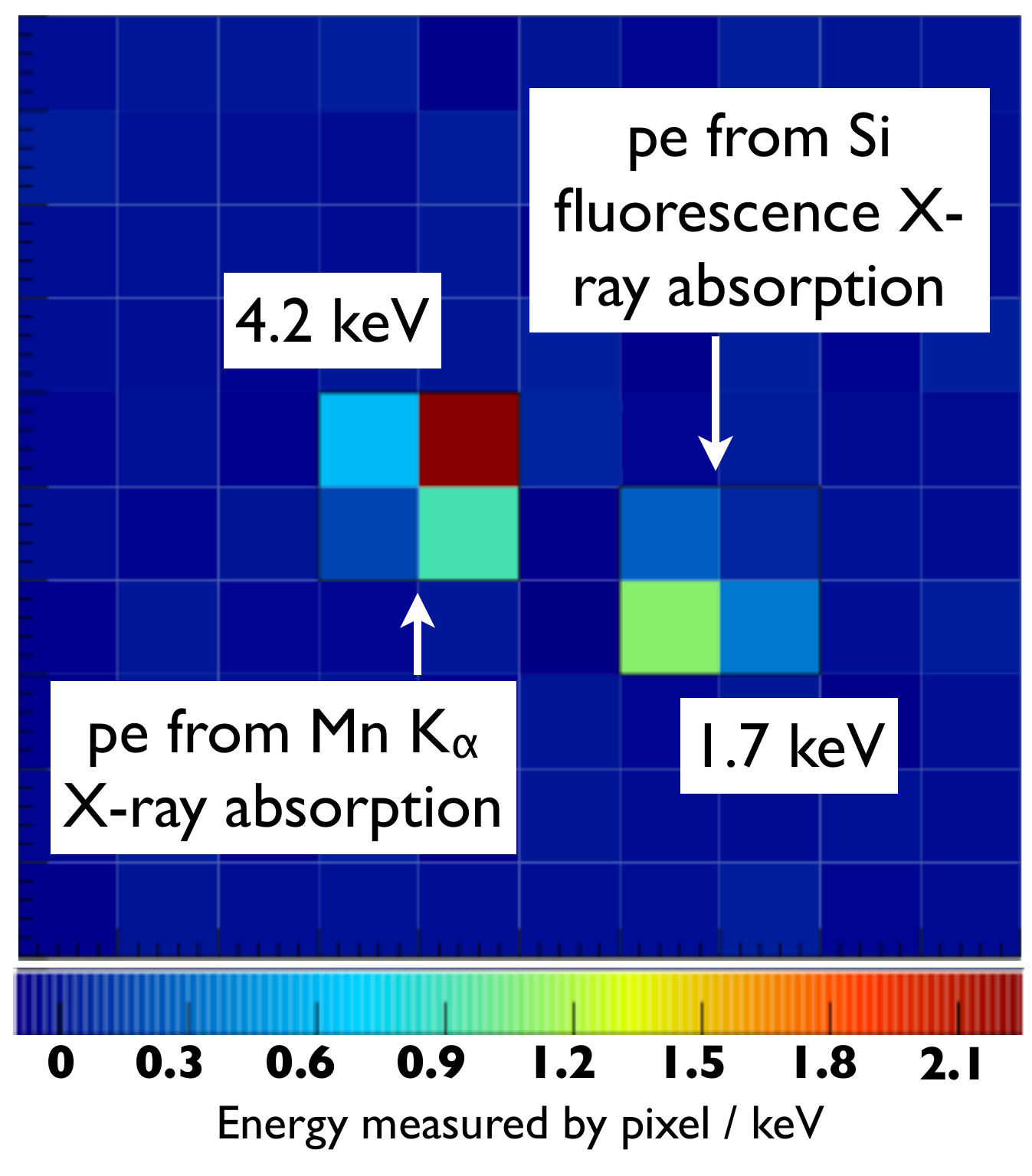}
\label{fig:escape_event}
}
\end{center}
\caption{a)~50$\times$50 pixel portion of a CCD image, taken when the detector was at ground level. Different kinds of particles are recognizable (see text). For better contrast, only pixels with deposited energy $>$0.1\,keV are represented in color. b)~Event with two nearby clusters detected after illuminating the CCD with a \ironfive\ source. The 1.7\,keV cluster is a photoelectron (pe) from the absorption of a Si fluorescence X-ray, emitted following photoelectric absorption of the incident 5.9\,keV Mn K$_\alpha$ X-ray in a nearby site. \label{fig:image_eg_}}
\end{figure}

The DAMIC CCDs were developed at Lawrence Berkeley National Laboratory MicroSystems Lab~\cite{1185186}, starting from an existing  design for the Dark Energy Survey (DES) camera (DECam)~\cite{Mclean:2012pka}. They feature a three-phase polysilicon gate structure with a buried p-channel. The pixel size is 15\,$\mu$m\,$\times$\,15\,\um\ and the active region of the detector is high-resistivity (10--20 k$\Omega$\,cm) n-type silicon hundreds of micrometers thick. The high-resistivity of the silicon allows for a low donor density in the substrate ($\sim$\powero{11}\,cm\up{-3}), which leads to fully depleted operation at reasonably low values of the applied bias voltage ($\sim$40\,V for a 675\,\um -thick CCD). The CCDs are typically 8 or 16 Mpixels, with surface areas of tens of square centimeters. Figure~\ref{fig:pixel_ccd_sketch} shows a cross-sectional diagram of a CCD pixel, together with an illustration of the WIMP detection principle.

When operated at full depletion, ionization produced in the active region will be drifted along the direction of the electric field ($z$-axis). The holes (charge carriers) will be collected and held near the p-n junction, less than 1\,\um\ below the gates. 
Due to the mobility of the charge carriers, the ionized charge will diffuse as it is drifted, with a spatial variance that is proportional to the carrier transit time. Charge produced by interactions closer to the back of the CCD will have longer transit times, leading to greater lateral diffusion. The lateral spread (width) of the charge recorded on the CCD $x$-$y$ plane may be used to reconstruct the $z$-coordinate of a point-like interaction~\cite{Chavarria201521}. For extended tracks, e.g. from electrons and muons, this effect leads to a greater width when the track is closer to the backside, which provides information on the track orientation.

The ionized charge is held at the CCD gates until the charge is read out. During readout, the charge is transferred vertically from pixel to pixel along each column by appropriate clocking of the 3-phase gates (``parallel clocks''), while higher frequency clocks (``serial clocks'') move the charge of the last row horizontally to a charge-to-voltage amplifier (``output node''). The inefficiency of charge transfer from pixel to pixel is as low as \powero{-6} and the readout noise for the charge collected in a pixel is $\sim$2\,$e^-$~\cite{Chavarria201521}. Since on average 3.6\,eV is required to ionize an electron in silicon, the readout noise corresponds to an uncertainty of $\sim$7\,eV in deposited energy. 
The image can then be reconstructed from the order in which the pixels were read out, and contains a two-dimensional stacked history (projected on the $x$-$y$ plane) of all particle interactions throughout the exposure. For rare-event searches, it is advantageous to take long exposures ($\sim$8~hours in DAMIC) in order to minimize the number of readouts, and thus the number of pixels above a given threshold due to noise fluctuations. Note that even with these long exposures the CCD dark current ($<$0.1\,$e^-$pix$^{-1}$day$^{-1}$ at the operating temperature of $\sim$140\,K)  contributes negligibly to the noise.

Figure~\ref{fig:image_eg} shows examples of particle tracks as recorded by a DAMIC CCD. Low energy electrons and nuclear recoils, whose physical track length is $<$15\,\um, produce ``diffusion limited'' clusters, where the spatial extension of the cluster is dominated by charge diffusion. Higher energy electrons, from either Compton scattering or $\beta$ decay, lead to extended tracks. Alpha particles in the bulk or from the back of the CCD produce large round structures due to the plasma effect~\cite{Estrada201190} (Section~\ref{sec:alphas}). Cosmic muons pierce through the CCD, leaving a straight track of minimum ionizing energy. The orientation of a muon track is immediately evident from its width, since the end-point of the track on the back of the CCD is much wider than the end-point at the front due to charge diffusion. 
In Figure~\ref{fig:escape_event}, we show a Mn X-ray interaction from a \ironfive\ source which further demonstrates the superb spatial resolution of the CCD. Usually, the full 5.9\,keV energy of the Mn K$_\alpha$ X-ray is deposited as a single cluster.  Rarely, a fluorescence X-ray is emitted following photo-electric absorption of the primary X-ray. The fluorescence X-ray may travel far enough within the CCD (a few attenuation lengths) to deposit its energy away from the first X-ray interaction, leading to two separate clusters.

The detectors present an excellent linearity and energy resolution (55\,eV RMS  at 5.9\,keV)~\cite{Chavarria201521} for electron-induced ionization, as measured with X-ray sources. 
The CCD response to $\alpha$ particles was calibrated with an \ame\ source; the energy scales from $\alpha$-induced and electron-induced ionization were found to be within  3\%, with an $\alpha$ energy resolution of 50\,keV at 5.5\,MeV~\cite{Estrada201190}.

\subsection{Setup at SNOLAB}
\label{sec:snolab}

\begin{figure}[t!]
\begin{center}
\includegraphics[width=\textwidth]{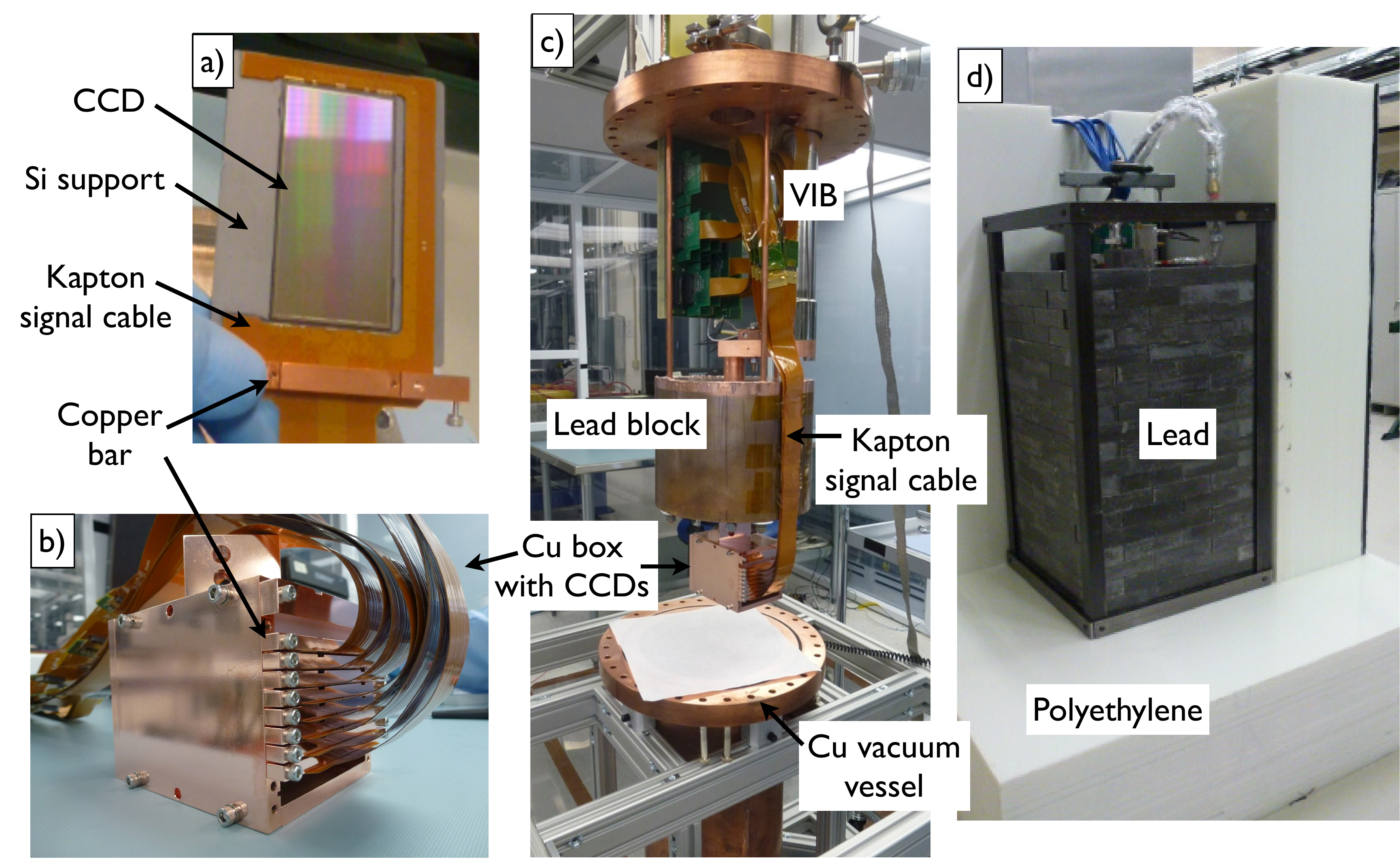}\\
\caption{a) A packaged DAMIC CCD. b) The copper box housing the CCDs. c) Components of the DAMIC setup, ready to be inserted in the vacuum vessel. d) The vessel inside the lead castle, during installation of the polyethylene shield.}
\label{detector_pictures}
\end{center}
\end{figure}

Most of the infrastructure for the DAMIC dark matter search is already installed in SNOLAB (Figure~\ref{detector_pictures}).
A packaged CCD (2k$\times$4k, 8~Mpixel, 500~\um -thick) is shown in Figure~\ref{detector_pictures}(a). The device is epoxied to a high-purity silicon support piece. The CCD clocks and output node signal travel on a Kapton cable, appropriately shaped for wire bonding. The cable is also glued to the silicon support. A copper bar facilitates the handling of the packaged CCD and its insertion into a slot of an electropolished copper box (Figure~\ref{detector_pictures}(b)). The box is cooled to $\sim$140\,K inside a copper vacuum vessel ($\sim$\powero{-6}\,mbar). 
An 18\,cm-thick lead block hanging from the vessel-flange shields the CCDs from radiation produced by the electronics card (Vacuum Interface Board, VIB), also located inside the vessel (Figure~\ref{detector_pictures}(c)). The CCDs are connected to the VIB through the Kapton flex cables, which run along the side of the lead block. The processed signals then proceed to the data acquisition electronic boards. The vacuum vessel is inserted in a lead castle (Figure~\ref{detector_pictures}(d)) which shields the CCDs from ambient $\gamma$-rays through at least 21\,cm of lead. The innermost inch of lead comes from an ancient Spanish galleon and has negligible \pbten\ content, strongly suppressing the background from bremsstrahlung $\gamma$s produced by \biten\ decays in the outer lead shield. A 42\,cm-thick polyethylene shielding is used to moderate and absorb environmental neutrons.

Three 2k$\times$4k, 8\,Mpixel CCDs were deployed in February 2014 as part of DAMIC R\&D efforts. Two of them were originally designed for astronomy, with a transparent indium-tin-oxide (ITO) coating deposited on the backside after thinning the CCD to 500\,\um\ . The third CCD was optimized for DAMIC by maximizing its mass (the CCD is un-thinned, 675\,\um-thick) and minimizing radioactive contamination (the ITO layer containing $\beta$-radioactive \indium\  is eliminated).
The 500-\um\ CCDs are inserted in adjacent slots of the copper box, with copper plates above and below. The 675-\um\ CCD is in a lower slot of the box, separated from the other CCDs by $\sim$1\,cm of copper.
DAMIC100 will consist of 18 CCDs, each of 4k$\times$4k pixels and 675\,\um\ thickness.

\subsection{CCD image reduction and data samples}
\label{sec:data}

Clusters of energy deposits are found in the acquired images with the following procedure. First, the pedestal of each pixel is calculated as its median value over the set of images. The pedestals are then subtracted from every pixel value in all images. Hot pixels or defects are identified as recurrent patterns over many images, and eliminated (``masked'') from the analysis ($>$95\% of the pixels were deemed good). Pixel clusters are selected as any group of adjacent pixels with signals greater than four times the RMS of the white noise in the image. The resulting clusters are considered candidates for particle interactions. Relevant variables (e.g. the total energy by summing over all pixel signals) are calculated for each cluster. For the studies presented in this paper, we required the cluster energy to be $>$1\,keV, which guarantees a negligible probability of accidental clusters from readout noise. Selection criteria specific to the different analyses will be described in Sections~\ref{sec:alpha} and~\ref{sec:spatcoinc}.

Two sets of data were collected, with different operating parameters for the CCD readout. The DAMIC readout employs a correlated double-sampling technique~\cite{janesick2001scientific}, where high-frequency noise is suppressed by measuring the voltage of the output node over relatively long intervals of time. The maximum pixel charge that can be recorded is limited by the dynamic range of the digitizer, since the pixel pedestal signal is proportional to the integration time. 
In the standard DAMIC readout, which is optimized for low noise, an integration time of 40\,$\mu$s per pixel corresponds to a maximum energy of $\sim$20\,keV. 
Radiogenic $\alpha$ particles have a range comparable to the pixel size, which leads to large localized charge deposits that can saturate the digitizer. Thus, a ``low-gain'' readout mode with an integration time of 0.6\,$\mu$s per pixel was used for runs dedicated to $\alpha$ spectroscopy.  Due to the shorter integration time, this set of data presents a higher pixel noise ($\sim$40\,eV RMS), which is still negligible for MeV $\alpha$  particles.

Table~\ref{tab:data} summarizes the data collected for this analysis. We used the low-gain data (0.6\,$\mu$s  integration time) to evaluate the \ura\ and \tho\ contamination in the CCD by $\alpha$ spectroscopy (Section~\ref{sec:alpha}). The rest of the data (40\,$\mu$s integration time) was taken after the installation of the ancient lead shield, and presents a tenfold decrease in the background of electron-like tracks. The rate of $\alpha$ particles did not change, since the CCDs and surrounding components inside the box, where the $\alpha$s must originate, were not touched. We used this data set to constrain the activities of \sitwo\ and \pbten\ in the bulk through a search of spatially correlated pairs of $\beta$ tracks (Section~\ref{sec:spatcoinc}). 

\begin{table}[t!]
\caption{Analyzed data sets. The background rate refers to electron-like tracks in the 675\,\um -thick CCD.
}
\centering
\begin{tabular}{|c|c|c|c|c|}\hline
Start date & End date & \underline{ Live-time$_{ }$}          & \underline{Pixel integration time}  & \underline{Background rate}  \\ 
		&		  &		day			     &			$\mu$s    		      &			g$^{-1}$\,d$^{-1}$	\\ \hline
2014/06/06 & 2014/07/07 & 28.7 & 0.6 & 45.9$\pm$0.7 \\ \hline
2014/07/11 & 2014/08/21  & 36.1 & 40 & 5.1$\pm$0.2 \\ \hline
2014/09/03 & 2014/09/29  & 20.7 & 40 & 4.8$\pm$0.3 \\ \hline
\end{tabular}
\label{tab:data}
\end{table}

\section{Limits on uranium and thorium contamination from $\alpha$ spectroscopy}
\label{sec:alpha}

\begin{figure}[t!]
\begin{center}
\includegraphics[width=0.8\textwidth]{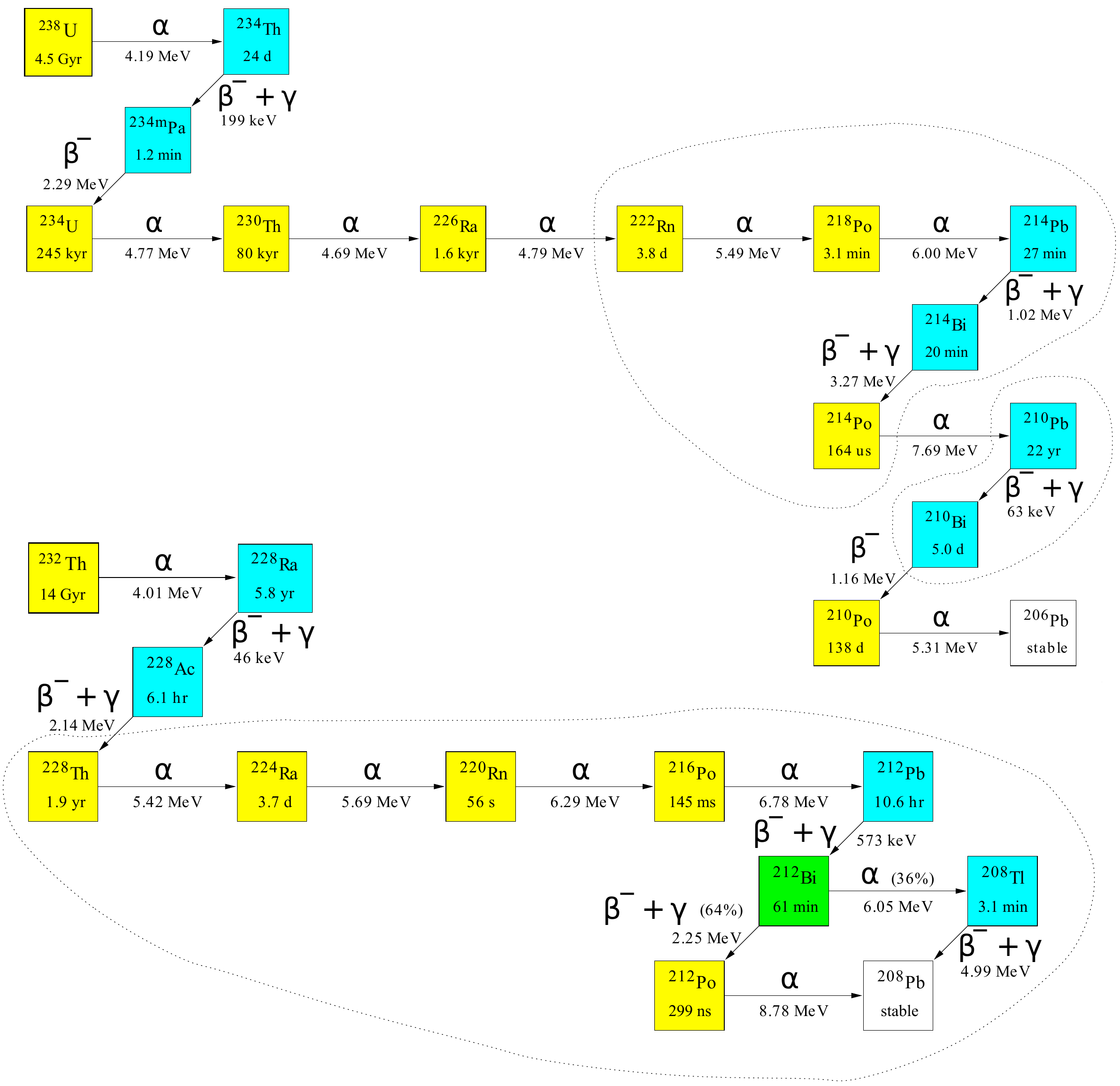}
\caption{\ura\ and \tho\ decay chains. Alpha ($\beta$) emitters are colored yellow (teal). For each isotope, $\alpha$ energies, $\beta$ $Q$-values and half-lives are given. Isotope sequences expected to be in secular equilibrium are grouped by dashed lines.} 
\label{fig:decay_chain}
\end{center}
\end{figure}

Uranium and thorium and their decay products (Figure~\ref{fig:decay_chain}) are ubiquitous in nature. They may be found within the bulk silicon of the CCD or the electrical elements of the device. They may also be present on the surface of the CCD, in particulates (e.g. dust) or as adsorbed radon daughters (mainly via deposition of \poeight\ from \radon\  decays in the air surrounding the device).

The CCD is $>$99.5\% electronic-grade silicon~\cite{vonAmmon198494} by mass. The remainder consists of light elements used in the oxidation, metallization and doping of the silicon to fabricate the semiconductor structures necessary for operation. Due to the extremely high chemical purity of the silicon, the substrate (photosensitive volume in Figure~\ref{fig:pixel_ccd_sketch}) can be doped with a donor density as low as $10^{11}$\,cm$^{-3}$~\cite{1185186}. Thus, even an unexpectedly large contamination of \ura\ or \tho\ with an atomic abundance $\sim$10\% of the donor density would only correspond to ppt (\powero{-12}) levels by mass.

The CCDs are consistently handled in ESD-safe clean rooms (class 1000 or better). DAMIC CCDs are packaged at Fermilab, with facilities developed for DECam, an astronomy camera whose requirements for the deposition of dust on the CCD surface are much more stringent than that of DAMIC. However, the clean rooms are not radon-free, and deposition of radon daughters, mostly on the front CCD surface, will occur during the packaging. In particular, after being glued to their support, CCDs are left for two days under an air column of a few centimeters in height to cure the epoxy. We estimate that this procedure may induce a residual surface activity of \pbten\ and its daughters of less than \powert{-3}{3}\,cm$^{-2}$\,d$^{-1}$, assuming a typical \radon\ activity for indoor air of 30\,Bq\,m$^{-3}$.

Many of the isotopes in the \ura\ and \tho\ decay chains (Figure~\ref{fig:decay_chain}) are $\alpha$ emitters, and can be efficiently identified by $\alpha$ spectroscopy.

\subsection{Characteristics and selection of $\alpha$-induced clusters}
\label{sec:alphas}

\begin{figure}[t!]
\begin{center}
\subfigure[Plasma $\alpha$]{
\includegraphics[width=0.25\textwidth]{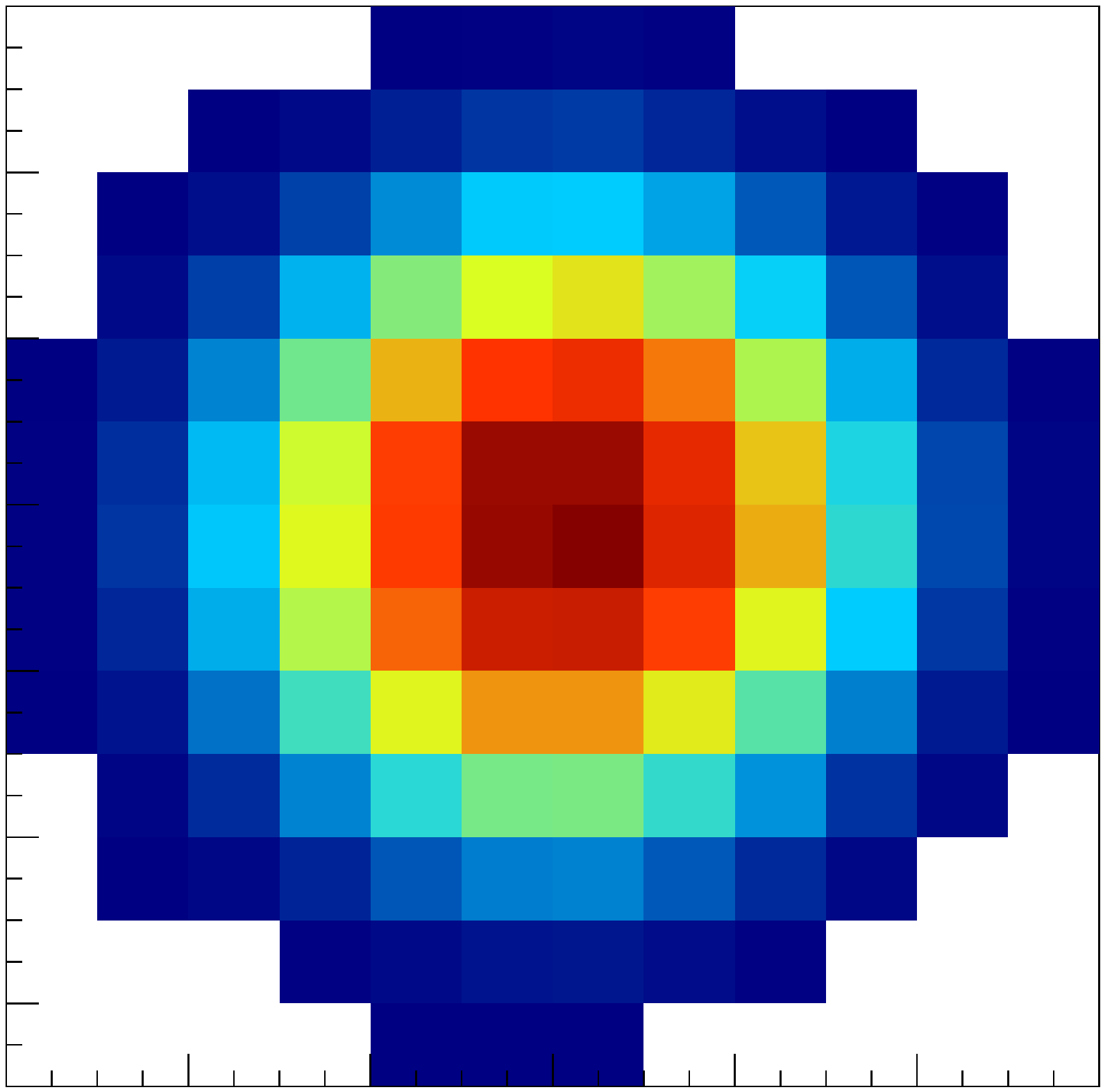}
\label{fig:alpha_plasma}
}
\hspace{1.5cm}
\subfigure[Bloomed $\alpha$]{
\includegraphics[width=0.11\textwidth]{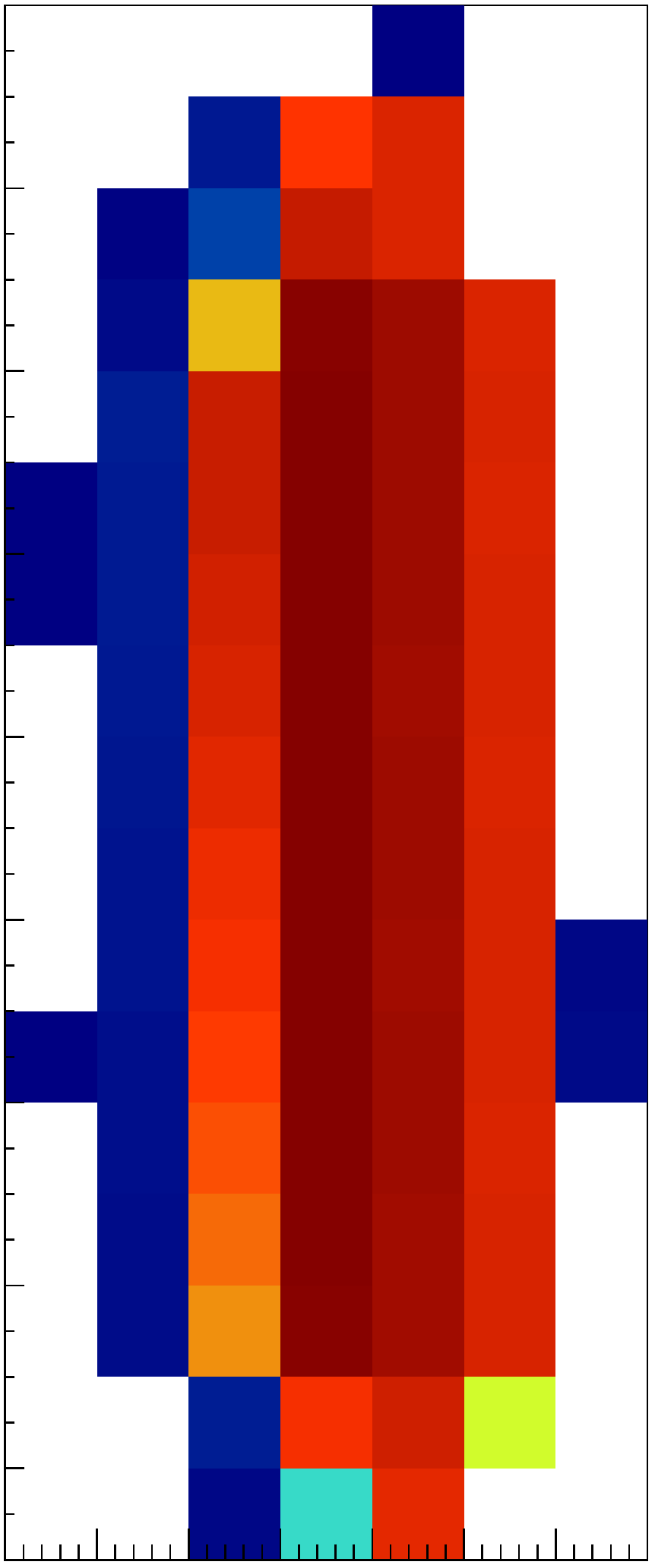}
\label{fig:alpha_bloom}
}
\end{center}
\caption{a) A highly-diffuse, round cluster due to plasma effect from an $\alpha$ particle originating in the bulk or the back of the CCD. b) An example of blooming, where an $\alpha$ particle originating in the front of the CCD produces a mostly vertical cluster. 
\label{fig:alpha_eg}}
\end{figure}

Radiogenic $\alpha$s lose most of their energy by ionization, creating a dense column of electron-hole pairs that satisfy the plasma condition~\cite{Estrada201190}. The local electric field within the plasma is much greater than the electric field applied across the substrate. For interactions deep in the substrate, where the electric field is only along $z$, the charge carriers diffuse laterally toward regions of lower charge concentration until the substrate electric field becomes dominant. Thus, $\alpha$s originating in the bulk or the back surface of the CCD lead to highly-diffuse, round clusters of hundreds of micrometers in diameter. On the other hand, $\alpha$ particles that strike the front of the CCD deposit their energy less than 20\,\um\ below the gates. The high-density charge cloud can easily overcome \textemdash\ and spillover \textemdash\ the barrier phases between vertical pixels, while it is harder for it to overcome the potential barrier of the vertical channel stops between columns. This phenomenon is known as blooming~\cite{janesick2001scientific}, and leads to mostly vertical clusters. 
Examples of ``plasma'' and ``bloomed'' $\alpha$s detected in DAMIC are shown in Figs.~\ref{fig:alpha_plasma}-~\ref{fig:alpha_bloom}.

\begin{figure}[t!]
\begin{center}
\subfigure[$\alpha$--$\beta$ selection]{
\includegraphics[width=0.475\textwidth]{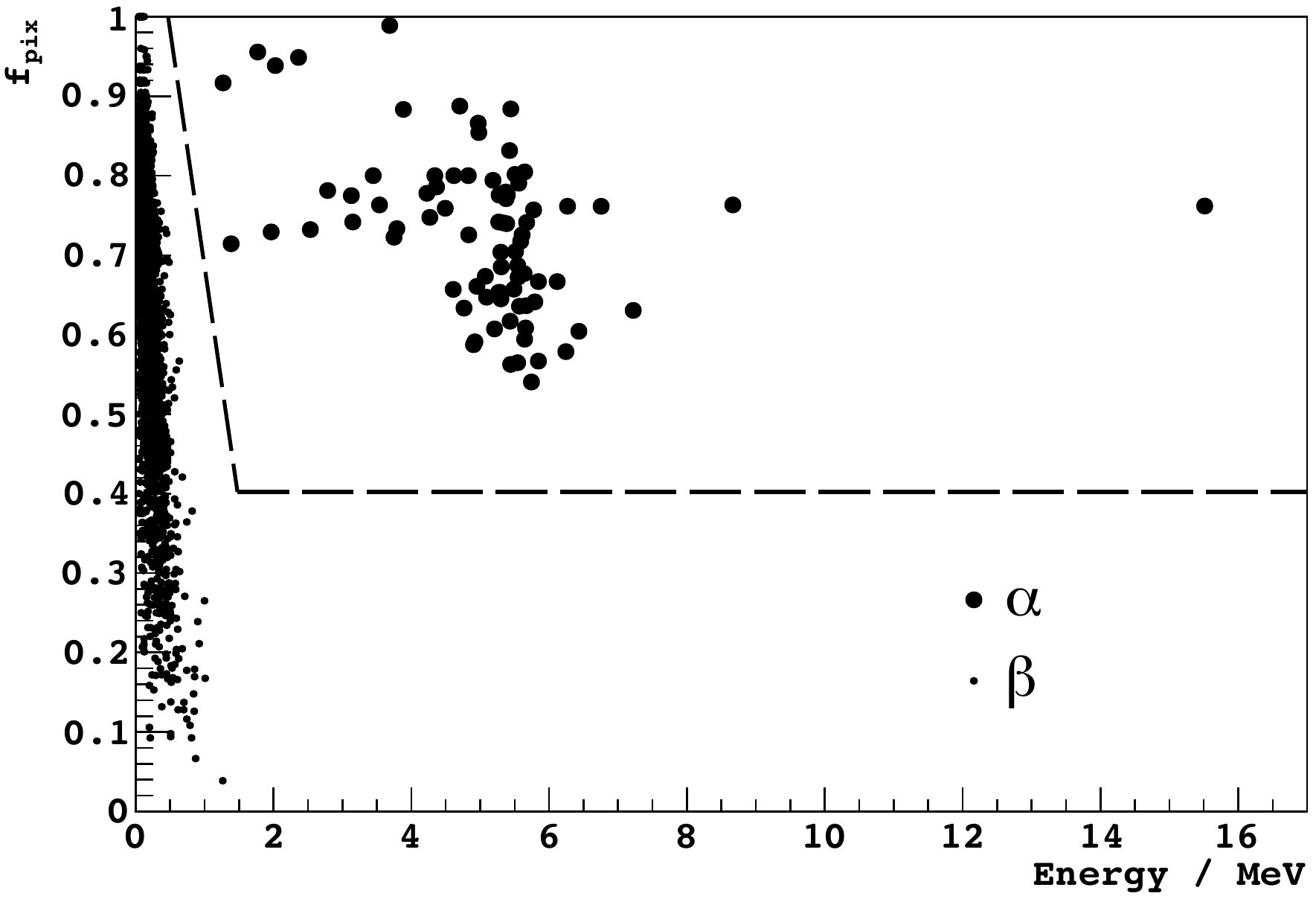}
\label{fig:alpha_beta_selection}
}
\subfigure[Plasma--bloomed $\alpha$ selection]{
\includegraphics[width=0.475\textwidth]{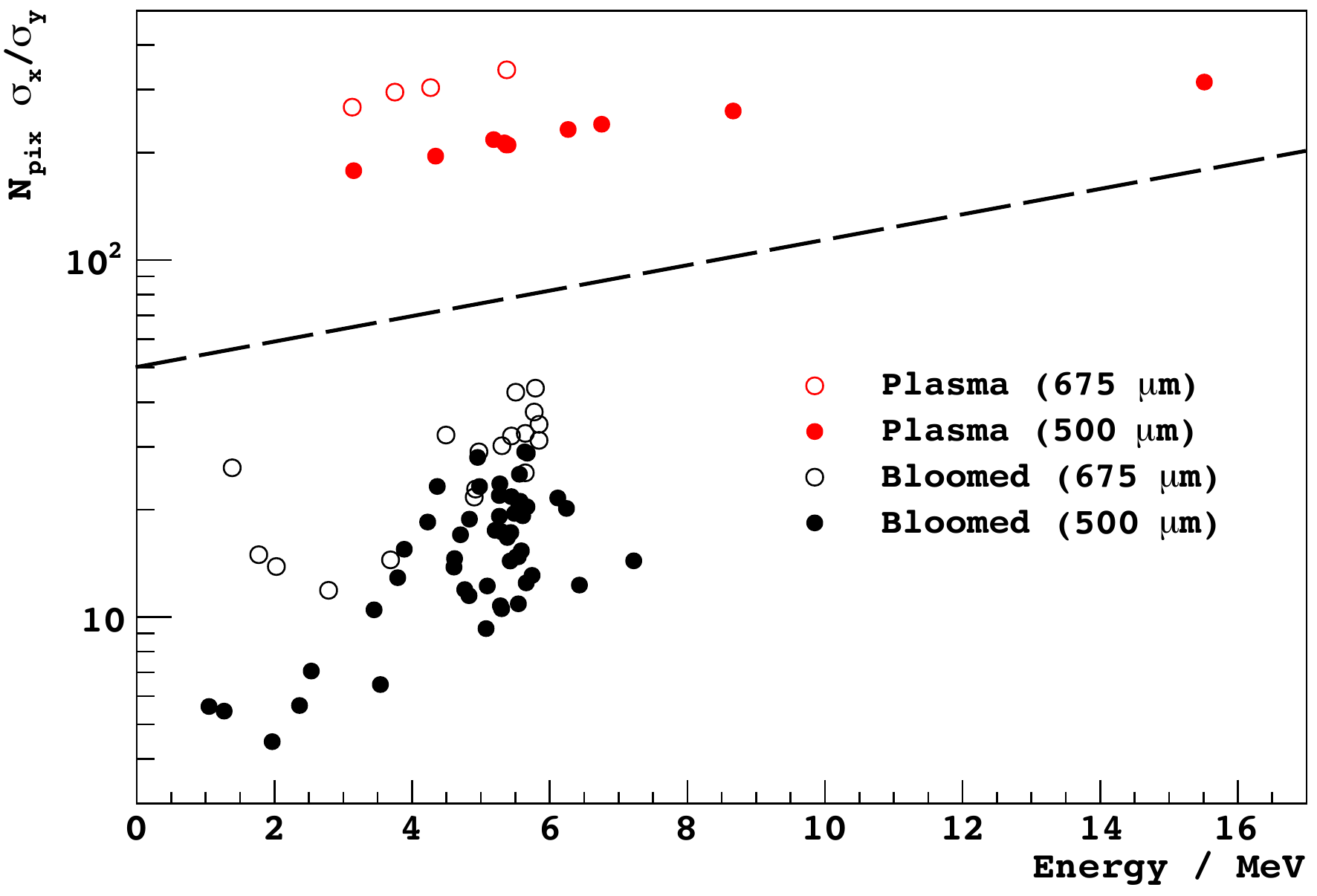}
\label{fig:bloom_plasma_selection}
}
\end{center}
\caption{Selection of $\alpha$ particles. a)~The fraction $f_{\rm{pix}}$ as a function of the cluster energy. Clusters in the region above the dashed line are selected as $\alpha$s. b)~The variable $N_{\rm{pix}}\,\sigma_{x}/\sigma_{y} $ as a function of the $\alpha$ energy. Plasma~(bloomed) $\alpha$s are indicated by red~(black) dots above~(below) the dashed line. Open~(closed) dots refer to clusters detected in the 675\,\um~(500\,\um)~-thick CCD. 
Clusters are more diffuse in the thicker CCD, resulting in larger $N_{\rm{pix}}$.     
\label{fig:alpha_selection}}
\end{figure}

Due to their distinctive features, simple criteria are sufficient to efficiently select and classify $\alpha$s. In the energy range of radiogenic $\alpha$s ($>$1\,MeV), electron tracks are long ($\sim$mm or more) and deposit their energy in extended ``worm''-like tracks over many CCD pixels. To differentiate electrons from $\alpha$s, we determine the smallest rectangular box that can contain a cluster, and compute the fraction of pixels, $f_{\rm{pix}}$, in this ``bound box'' which are part of the cluster. For small, symmetric clusters (i.e. $\alpha$s) $f_{\rm{pix}}$ is large ($\sim$$\pi/4$ for a round cluster). For the long and irregularly shaped worms characteristic of electrons, $f_{\rm{pix}}$ is small and decreases with increasing electron energy. Figure~\ref{fig:alpha_beta_selection} shows the successful separation between $\beta$s and $\alpha$s in the low-gain data set according to this variable.

To further distinguish plasma from bloomed $\alpha$s, we calculate the spatial RMS $\sigma_{x,y}$ of the cluster as the signal-weighted RMS value of the pixels' $x$,$y$ coordinate. Plasma $\alpha$s present a round-shaped cluster, with $\sigma_{x}/\sigma_{y} \sim 1$, while bloomed $\alpha$s are generally longer along the $y$ axis, giving $\sigma_{x}/\sigma_{y} < 1$. In addition, the diffused clusters from plasma $\alpha$s have more pixels ($N_{\rm{pix}} $) than bloomed $\alpha$s. We used the variable $N_{\rm{pix}}\,\sigma_{x}/\sigma_{y} $ to separate 
plasma from bloomed $\alpha$s (Figure~\ref{fig:bloom_plasma_selection}).

\subsection{Limits on radioactive contamination from $\alpha$ analysis}
\label{sec:alpha_estimates}

The measured rate of $\alpha$s in the three installed CCDs is reported in Table~\ref{tab:alphas}. 
  
\begin{table}[t!]
\caption{CCD physical properties and rate of observed $\alpha$s for the three CCDs installed at SNOLAB.}
\centering
\begin{tabular}{|c|c|c|c|c|}\hline
CCD 			& Mass / g & Area / cm$^{2}$ & Bloomed rate / d$^{-1}$ 	& Plasma rate / d$^{-1}$  \\ \hline
500\,\um\ top 	& 2.2 & 19 & 0.87$\pm$0.17 				&  0.21$\pm$0.09			  \\ \hline
500\,\um\ bottom 	& 2.2 & 19 &0.87$\pm$0.17 				&  0.14$\pm$0.07  		  \\ \hline
675\,\um\ 		& 2.9 & 19 &0.63$\pm$0.15  			&  0.14$\pm$0.07	   	 	 \\ \hline
\hline
Average			& 2.4 & 19 & 0.79$\pm$0.10				&  0.16$\pm$	0.04			  \\ \hline
\end{tabular}
\label{tab:alphas}
\end{table}
  
Most of the bloomed $\alpha$s are clustered around the characteristic energy of  \poten\ decay (5.3\,MeV, Figure~\ref{fig:bloom_plasma_selection}), which may be present as residual surface contamination following exposure to \radon. Also, a significant number of $\alpha$s have energies $<$4\,MeV, lower than any $\alpha$s from the \ura\ and \tho\ chains (Figure~\ref{fig:decay_chain}). Most likely, these are $\alpha$ particles which lose some energy before reaching the active region of the device, and originate from surface contamination of the CCD or nearby materials.
 If we conservatively assume that all bloomed $\alpha$s with energies $<$6\,MeV are due to \poten\ decays from \pbten\ contamination on the front surface of the CCD, we obtain an activity of 0.078$\pm$0.010\,cm$^{-2}$\,d$^{-1}$. This is twenty times larger than the upper limit of \powert{-3}{3}\,cm$^{-2}$\,d$^{-1}$ previously estimated from the exposure to \radon\ during packaging. However, a significant number of the observed $\alpha$s are likely to originate from the surfaces facing the CCD, i.e. the copper plates above the 675\,\um\ and the top 500\,\um\ CCD, and the silicon support piece above the bottom 500\,\um\ CCD. Unfortunately, contributions from all these different surfaces cannot be disentangled with the available data. Similar considerations can be applied to plasma $\alpha$s, which should release their full energy if occurring in the bulk. Many of the plasma $\alpha$s have energies lower than those from \ura\ and \tho\ chains, suggesting surface contamination. Taking again the conservative assumption that all plasma $\alpha$s with energies $<$6\,MeV originate from \poten\ back-surface contamination, we obtain an activity of 0.012$\pm$0.004\,cm$^{-2}$\,d$^{-1}$.

Spectroscopy of plasma $\alpha$s can be used to establish limits on  \pbten\ , \ura\ and \tho\ contamination in the bulk of the CCD. 
Four plasma $\alpha$s whose energies are consistent with \poten\ were observed (Figure~\ref{fig:bloom_plasma_selection}). One of them cannot be \poten, as it coincides spatially with two higher energy $\alpha$s recorded in different CCD exposures, and is therefore likely part of a decay sequence (Section~\ref{sec:alpha_seq}). When interpreting the other three as bulk contamination of \poten\ (or \pbten), an upper limit of $<$37\,kg$^{-1}$\,d$^{-1}$ (95\% CL) is derived. In the \ura\ chain, the isotopes \urafour, $^{230}$Th and \radium\ decay by emission of $\alpha$s with energies 4.7--4.8\,MeV (Figure~\ref{fig:decay_chain}). Since the isotopes' lifetimes are much longer than the CCD exposure time, their decays are expected to be uncorrelated. No plasma $\alpha$s were observed in the 4.5--5.0\,MeV energy range, and an upper limit on the \ura\ contamination of $<$5\,kg$^{-1}$\,d$^{-1}$ or $<$4\,ppt (95\% CL) is correspondingly derived (secular equilibrium of the isotopes with \ura\ was assumed).
In the \tho\ chain, the timescale of the short-lived decay sequence of $^{224}$Ra\,--\thoron\,--$^{216}$Po is $\sim$1\,min, which is much smaller than the CCD exposure time. Thus, such a decay sequence would result in a single cluster when occurring in the bulk, with a total of 18.8\,MeV of energy deposited by the pile-up of the three decays. No cluster with energy $>$16\,MeV is observed, which results in an upper limit of $<$15\,kg$^{-1}$\,d$^{-1}$ or $<$43\,ppt (95\% CL) on \tho\ contamination in the CCD bulk.

\subsection{Observation of spatially correlated $\alpha$ decay sequences}
\label{sec:alpha_seq}

\begin{figure}[t!]
\begin{center}
\subfigure[Triple $\alpha$ sequence]{
\includegraphics[width=0.65\textwidth]{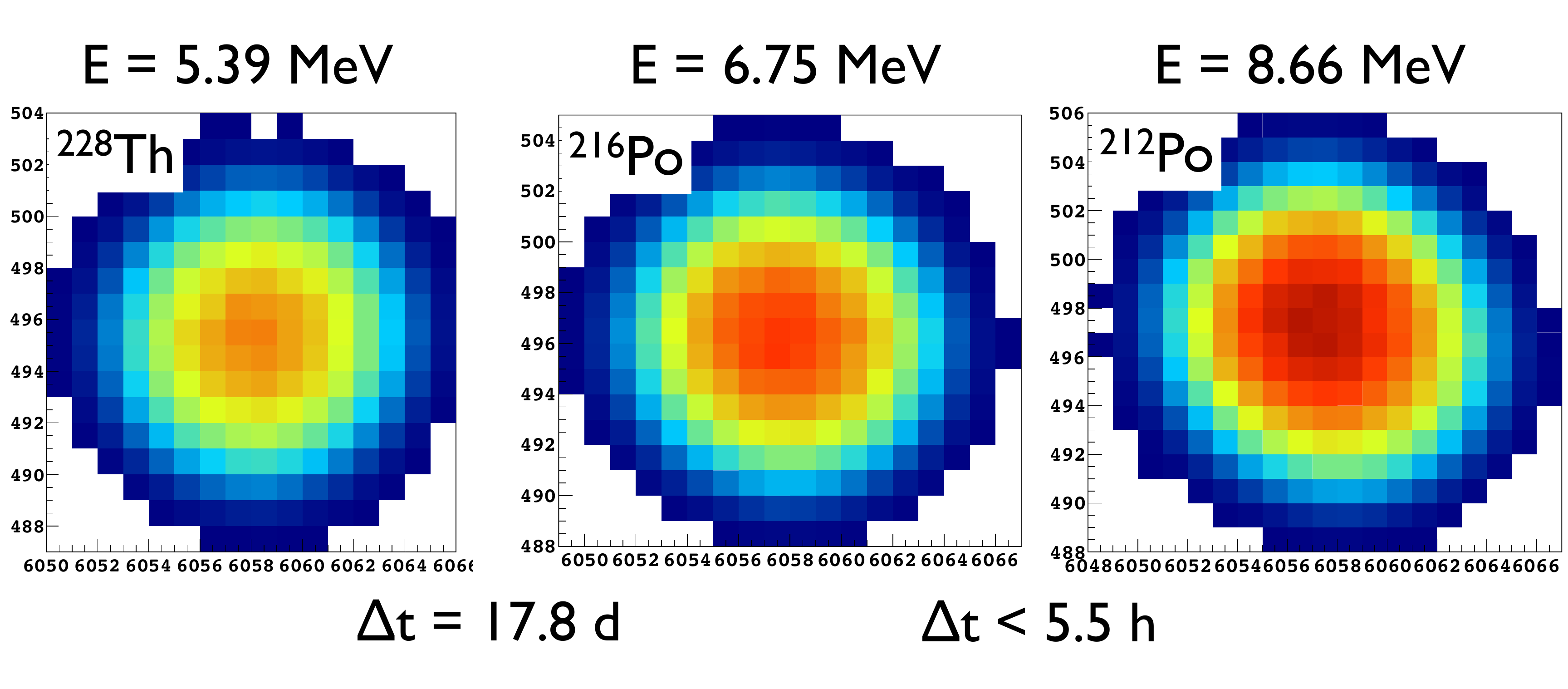}
\label{fig:triple_alpha}
}
\hspace{1.cm}
\subfigure[$\alpha$--$\beta$ coincidence]{
\includegraphics[width=0.23\textwidth]{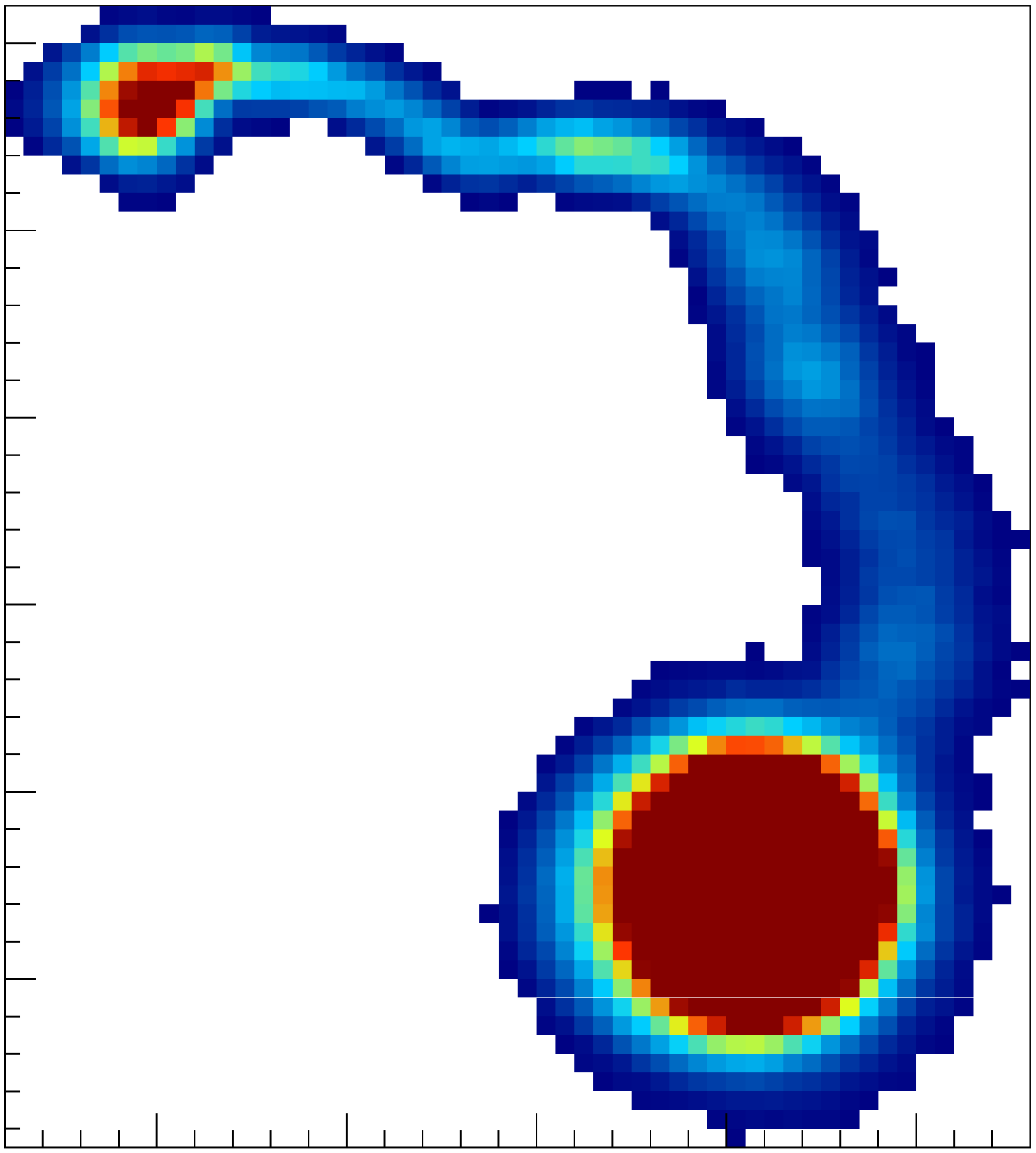}
\label{fig:alpha_beta}
}
\end{center}
\caption{a) Three $\alpha$ particles detected in different CCD images at the same $x$-$y$ position. Their energies and the time separation between images are consistent with a sequence from a \tho\ decay chain. b) A peculiar cluster found in a single image, consistent with a plasma $\alpha$ and a $\beta$ track originating from the same CCD position. This may happen for a radioactive decay sequence occurring within the 8-hour exposure time of an image. 
\label{fig:th_sequence}}
\end{figure}

We detected four plasma $\alpha$s with energies $>$5.5\,MeV (Figure~\ref{fig:bloom_plasma_selection}), which cannot be accounted for by \poten\ decay. They all occur in the top 500-\um\ CCD. Two of them, along with an $\alpha$ close in energy to the \poten\ line, have their centroids within 1\,pixel on the CCD $x$-$y$ plane (Figure~\ref{fig:triple_alpha}). Since the accidental probability of this occurence is negligible, these three strongly spatially correlated events must have a common origin. A likely explanation is that we have observed a decay sequence starting with a $^{228}$Th nucleus located in the thin (60\,nm) ITO layer covering the backside of the CCD.  The energies of the $\alpha$s and the time separation between decays are consistent with those from the decays of $^{228}$Th, $^{216}$Po and $^{212}$Po (Figure~\ref{fig:decay_chain}). The two other $\alpha$s of the sequence, $^{224}$Ra and $^{220}$Rn, must have been emitted away from the CCD and thus gone undetected.  Note that the recoiling nuclei corresponding to the $^{224}$Ra and $^{220}$Rn decays may penetrate the device, but their deposited energy would be too small to be observable on top of the $^{216}$Po $\alpha$ once the detector resolution is considered. Likewise, the $\beta$s from $^{212}$Pb and $^{212}$Bi must have been emitted away from the CCD or have deposited $<$100--200\,keV in the CCD. The observation of this single decay sequence is compatible with $\sim$100\,ppb of \tho\ contamination in the ITO~\cite{Groom:2002fk}.

The highest energy plasma $\alpha$ (15.3\,MeV) must be due to pile-up, either of two $\alpha$s (e.g. $^{216}$Po and $^{212}$Po) or $\alpha$s and $\beta$s with energies $<$100--200\,keV (so that the beta track is too short to protrude outside of the $\alpha$ cluster). Likewise, the remaining 6.2\,MeV $\alpha$ could be due to either a single $\alpha$ (e.g. \thoron) or a less energetic $\alpha$ piled-up with a $<$100-200\,keV $\beta$. In either case, these decays are unlikely to have occurred in the bulk of the CCD, as there are no isolated decays or short sequences of decays ($\lesssim$hours) corresponding to these energies in the \ura\ or \tho\ decay chains (Figure~\ref{fig:decay_chain}).

A spatial coincidence search between $\alpha$s and $\beta$s is limited by the large number of accidentals from background, as the $\beta$ rate is $\sim$100 times greater than the $\alpha$ rate in the low-gain data set. In one of the latter data sets we have observed a coincidence between an $\alpha$ and a $\beta$ within the same exposure, shown in Figure~\ref{fig:alpha_beta}. As the energy of the $\alpha$ cannot be measured due to digitizer saturation, this particular decay sequence cannot be identified.

\section{Limits on \sitwo\ and \pbten\  contamination from $\beta$ decay sequences}
\label{sec:spatcoinc}

We have performed a search for decay sequences of two $\beta$ tracks to identify radioactive contamination from \sitwo\ and \pbten\ and their daughters, whose $\beta$ spectra extend to the lowest energies and could represent a significant background in the region of interest for the WIMP search. These isotopes do not emit $\alpha$ or penetrating $\gamma$ radiation, and their decay rates are significant for extremely low atomic abundances due to their 10--100\,y half-lives, making conventional screening methods ineffective in determining their presence at the low levels necessary for a WIMP search.

\sitwo\ is produced by cosmic ray spallation of argon in the atmosphere, and then transported to the Earth's surface, mainly by rain and snow. Detector-grade silicon is obtained through a chemical process starting from natural silica. Therefore, the \sitwo\ content of a silicon detector should be close to its natural abundance in the raw silica. Spectral measurements of radioactive background in silicon detectors suggest a rate of \sitwo\ at the level of hundreds of decays per kg\,day~\cite{PhysRevLett.65.1305}. \sitwo\ leads to the following decay sequence:

\begin{align}
\mbox{\sitwo} &\longrightarrow \mbox{\ptwo} +\beta^-  ~\rm{with } ~\tau_{1/2} = \rm{150\,y, ~Q-value=227\,keV}\\
\mbox{\ptwo} &\longrightarrow \mbox{$^{32}$S} +\beta^- ~ \rm{with} ~\tau_{1/2} = \rm{14\,d, ~Q-value = 1.71\,MeV}
\end{align}

\pbten\ is a member of the \ura\ decay chain (Figure~\ref{fig:decay_chain}) and is often found out of secular equilibrium, as chemical processes in the manufacture of materials separate it from other \ura\ daughters. It may also remain as a long-term surface contaminant following exposure to environmental \radon\ (Section~\ref{sec:alpha}). \pbten\ leads to the following decay sequence:

\begin{align}
\mbox{\pbten} &\longrightarrow \mbox{\biten} +\beta^- + \mbox{IC}\mbox{ (80\%) } / ~\gamma \mbox{ (4\%)}  ~\rm{with } ~\tau_{1/2} = \rm{22\,y, ~Q-value=63.5\,keV} \label{eq:pbten}\\
\mbox{\biten} &\longrightarrow \mbox{\poten} +\beta^- ~ \rm{with} ~\tau_{1/2} = \rm{5\,d, ~Q-value=1.16\,MeV}
\end{align}

The \pbten\ nucleus decays 84\% of the time into an excited state of \biten\, which promptly releases its 46.5\,keV of energy by internal conversion of an atomic electron (IC) in 80\% of the decays or by emission of a $\gamma$-ray in 4\% of the decays. \poten\ is itself radioactive and decays by $\alpha$ emission. The possible contamination from \poten\ in the CCD has been discussed in Section~\ref{sec:alpha_estimates}.

\begin{figure}[t!]
\begin{center}
	\includegraphics[width=0.7\textwidth]{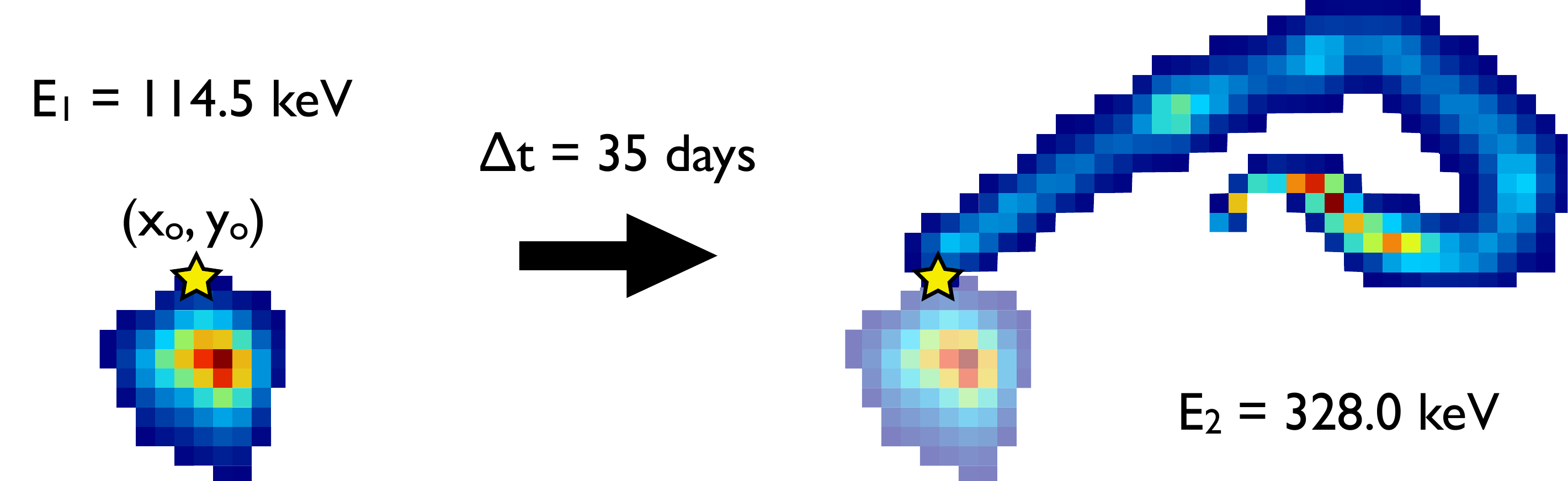}\\
	\caption{Candidate $\beta$ decay sequence found in data. The first cluster was detected in an image taken on 2014/08/05 and deposited 114.5\,keV of energy. A second cluster, with energy 328.0\,keV, was observed in an image taken 35 days later. Both tracks appear to originate from the same point (yellow star) in the CCD $x$-$y$ plane. 
	\label{fig:si32_candidate}}
\end{center}
\end{figure}

The intermediate nuclei, \ptwo\ and \biten, are expected to remain in the same lattice site as their parent nuclei and throughout their lifetimes. Therefore, the $\beta$s produced by each decay pair should originate from the same pixel (out of \powert{6}{8}) on the $x$-$y$ plane of the CCD. Through a search for electron-like tracks starting from the same spatial position, individual \sitwo\,--\ptwo\ and \pbten\,--\biten\ decay sequences can be selected with high efficiency. We performed this search with the lowest background data set (Table~\ref{tab:data}) in the 675\,\um\ CCD. Given the background level ($\sim$10\,electrons per day in a CCD), the number of accidental coincidences among uncorrelated tracks are small for periods of time comparable to the half-lives of \ptwo\ and \biten. A candidate decay sequence found in the data is shown in Figure~\ref{fig:si32_candidate} to illustrate the search strategy.

\subsection{Search procedures for spatially correlated $\beta$ decay sequences}
\label{sec:procedure}

\begin{figure}[t!]
\begin{center}
	\includegraphics[width=\textwidth]{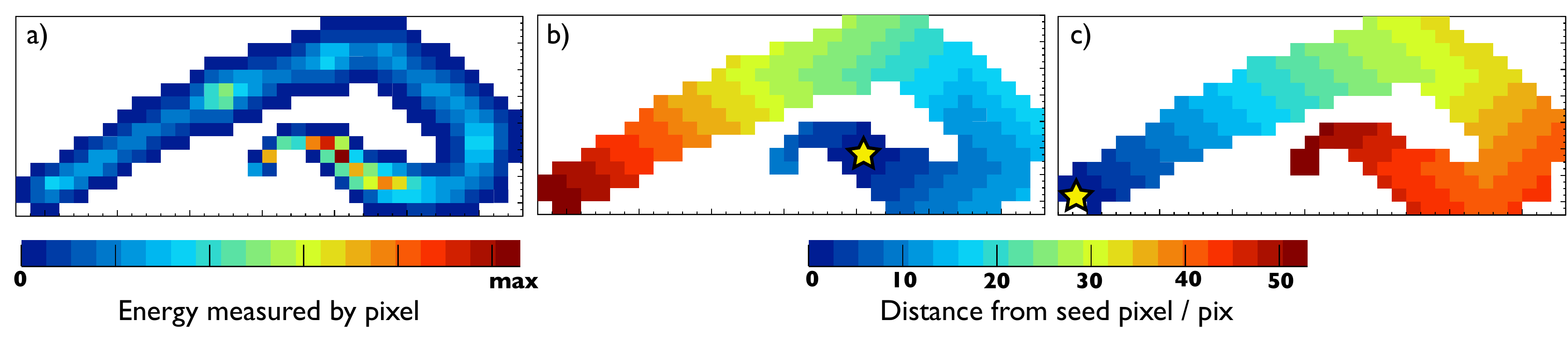}\\
	\caption{Algorithm to find the end-points of a cluster. a)~The pixel with maximum signal is chosen as seed point. b)~The distance of each pixel in the cluster to the seed point (star) is computed. c) The pixel with the largest distance is chosen as the first end-point (star). Distances to the first end-point are calculated, and the pixel with the largest distance is taken as the second end-point (the reddest pixel). 
	\label{fig:track_procedure}}
\end{center}
\end{figure}

The first step in the search for decay sequences is to find the end-points of the $\beta$ tracks. The procedure is illustrated in Figure~\ref{fig:track_procedure}.
First, we find the pixel with the maximum signal in the cluster, and we use it as a seed point. Then, for every pixel of the cluster we compute the length of the shortest path to the seed point, where the path is taken only along pixels that are included in the cluster. We refer to this as the ``distance'' from the seed point. The pixel with the greatest distance is taken as the first end-point of the track. Finally, we recompute the distance of every pixel from the first end-point, and take the pixel with the largest distance as the second end-point of the cluster. 

To find a $\beta$ decay sequence, we calculate the distance from the end-points of every $\beta$ cluster in an image to the end-points of every $\beta$ cluster in later images. Thus, for every pair of clusters we have four distances corresponding to each end-point combination. The minimum of these distances is defined as the ``cluster distance.'' The pair is considered a candidate for a decay sequence if the cluster distance is smaller than 20 pixels and the clusters have at least one pixel in common. We refer to the cluster in the earlier (later) image as the ``first'' (``second'') cluster.

To reduce the number of accidental pairs, we impose additional criteria on the energy of the clusters and their time separation. For the \sitwo\,--\ptwo\ sequence search, we require the energy of the first cluster to be $<$230\,keV and the energy of the second cluster to be $<$1.8\,MeV. For the \pbten\,--\biten\ sequence search, we require the energy of the first cluster to be in the range 30--65\,keV, which mostly includes the 80\% of decays with an electron from internal conversion (Eq.~\ref{eq:pbten}). Clusters from IC will be diffusion limited, as they will be constituted by a cascade of $\beta$, conversion and Auger electrons with energies $<$30\,keV. The energy of the second cluster is restricted to be $<$1.2\,MeV. 

Lastly, we require the time separation between the clusters of each pair, $\Delta t_{\rm{pair}}$, to be less than five half-lives of the daughter nuclei. This corresponds to 70 (25) days for the \sitwo\,--\ptwo\ (\pbten\,--\biten) decay sequence search. 

\subsection{Pair selection efficiency}
\label{sec:efficiency}

The efficiency of the pair selection described in Section~\ref{sec:procedure} was estimated by Monte Carlo simulations. We used the MCNPX5~\cite{mcnp} program with full electron tracking to simulate $\beta$ particle interactions in a rectangular block of silicon with the same dimensions of a DAMIC CCD. The energy deposited by a particle is recorded in a mesh with cells of the size of a CCD pixel. The average $z$-coordinate of the simulated particle track within each cell is also recorded. Then, a realistic resolution and diffusion model is applied to the number of ionized charge carriers from the position at which they were produced along the electron's path, and a simulated pixel cluster on the $x$-$y$ plane of the CCD is obtained. For the purpose of the decay sequence search, the origin of the $\beta$ particle was randomly generated in the silicon volume of the CCD to simulate a decay occurring in the bulk. 
 
\begin{figure}[t!]
\begin{center}
\subfigure[Simulated sample search]{
\includegraphics[width=0.475\textwidth]{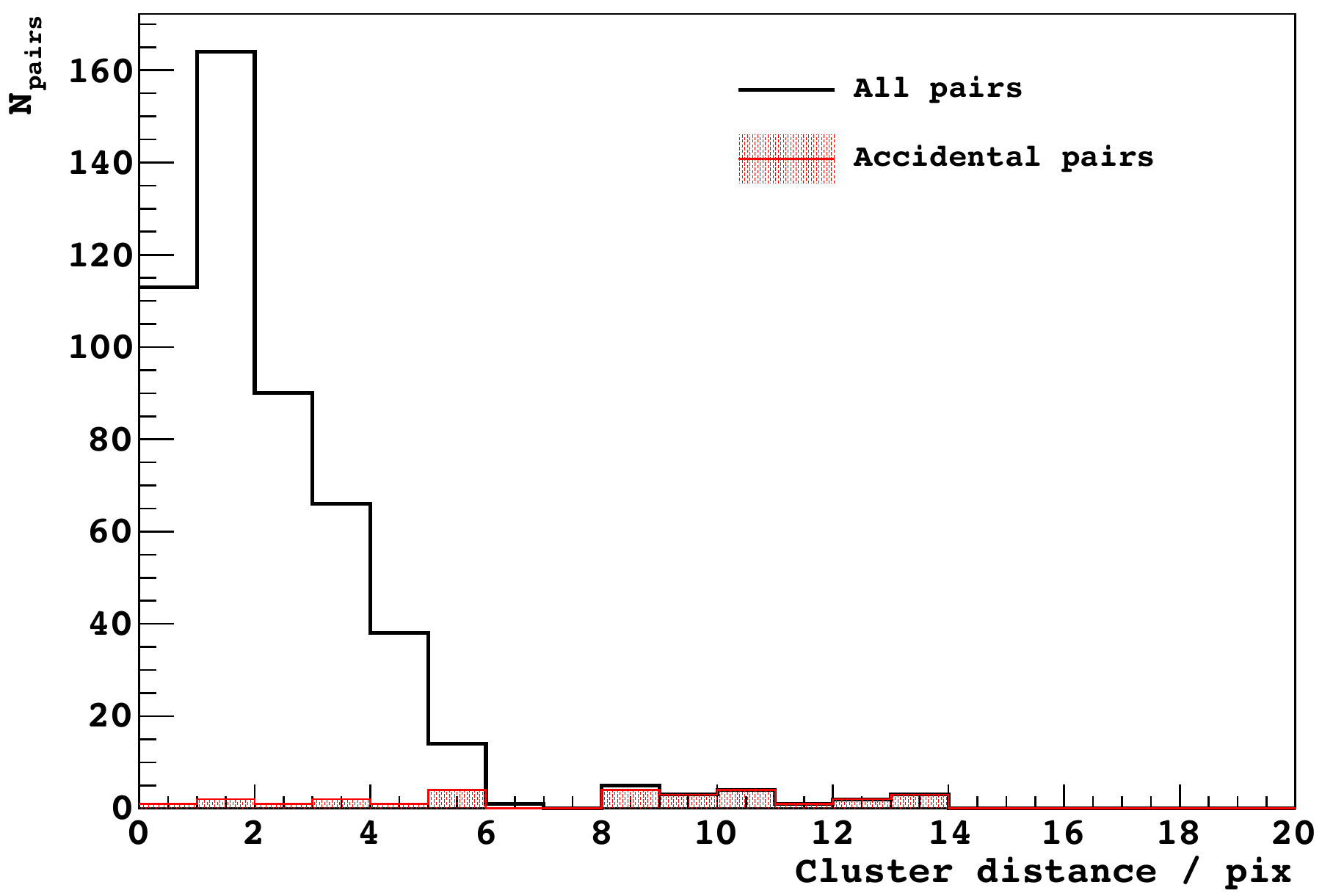}
\label{fig:distance_sim}
}
\subfigure[$\epsilon_{\rm{time}}$ for \sitwo\,--\ptwo]{
\includegraphics[width=0.4675\textwidth]{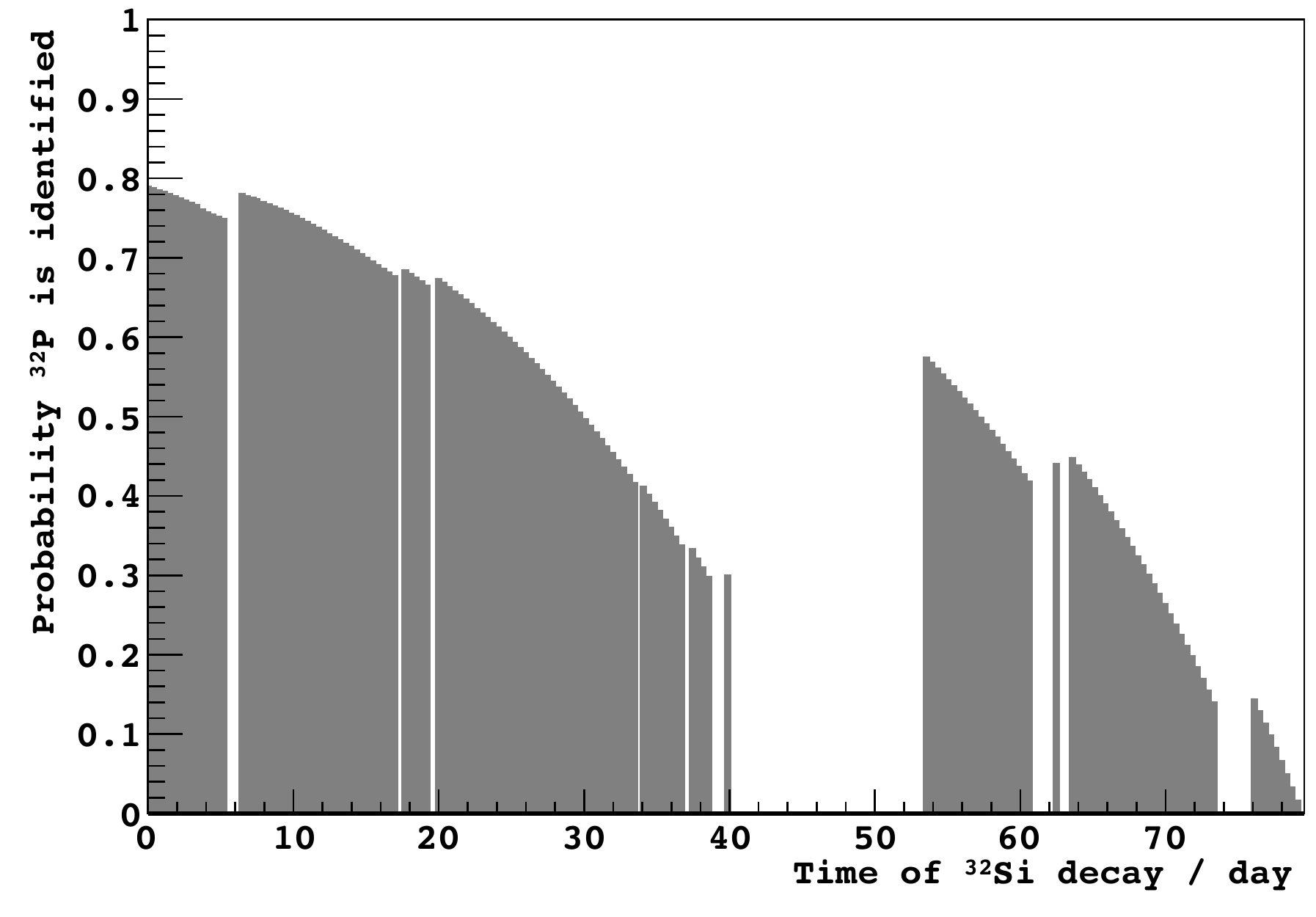}
\label{fig:si32_efficiency}
}
\end{center}
\caption{a)~Cluster distance distribution (black histogram) for the \sitwo\,--\ptwo\ candidate pairs found in the simulated sample. Accidental pairs are represented by the red shaded histogram.   b)~Probability for the \ptwo\ decay to be identified within the real-time of the data taking period, as a function of the time of decay of the parent \sitwo. The probability is zero when data was not acquired. 
\label{fig:distance_dist}}
\end{figure}

To properly include the measured readout noise in the analysis, we used data ``blanks.'' These are zero-length exposures read out immediately after every data exposure, which feature true readout noise patterns but no physical tracks. There are 159 blanks taken during the decay sequence search period. In each of them, we introduced the simulated clusters of five $\beta$s from \sitwo\ to approximately reproduce the rate of electron-like tracks measured in data. For each simulated \sitwo\ decay, a \ptwo\ decay was generated from the same location in a later image, distributed in time according to the half-life of \ptwo. With this method, 496 decay pairs were introduced in the sample of blank images. We used the standard CCD image reduction (Section~\ref{sec:data}) to reconstruct clusters. To properly account for inefficiencies of the CCD, bad pixels found in data were also masked in the blanks. Then, the pair selection procedure was applied to this set of simulated images. We found 504 candidate pairs ($N_{\rm{pairs}}$) with a cluster distance distribution shown in Figure~\ref{fig:distance_sim}. As most of the pairs have a cluster distance $<$7 pixels, we adopt this additional criterion for the pair selection.

First, we estimated the efficiency $\epsilon_{\rm{pair}}$ of requiring a cluster distance $<$7 pixels and cluster energies consistent with a \sitwo\,--\ptwo\  decay sequence (Section~\ref{sec:procedure}). Of the 486 pairs with cluster distance $<$7\,pixels, 475 are true pairs. Thus, the pair selection procedure is highly efficient in recovering decay sequences ($\epsilon_{\rm{pair}}$$=$475/496=95.8\%), while keeping background at a reasonable level.  The inefficiency is mostly due to pairs for which one or both tracks were not properly reconstructed due to masked pixels (16 out of 21 lost pairs). 

Second, we determined the efficiency $\epsilon_{\rm{time}}$ to select a \sitwo\,--\ptwo\ decay sequence with pairs separated in time by less than 70 days (Section~\ref{sec:procedure}) when the parent \sitwo\ decays during the live-time of our data set. In Figure~\ref{fig:si32_efficiency}, the probability for a \ptwo\ decay to be identified within the live-time of the data taking period is calculated as a function of the time of decay of the parent \sitwo. Integrating over the live-time of the data set gives an average $\epsilon_{\rm{time}}=$51.3\%.

An analogous simulation study was performed for the \pbten\,--\biten\ decay sequence. In this case, $\epsilon_{\rm{pair}}$ is mainly determined by the requirement for the energy of the first cluster to be in the range 30--65\,keV (Section~\ref{sec:procedure}). To estimate this efficiency, we have considered all possible processes following  \pbten\ decay, including the cases where some of the energy is radiated in $\gamma$-rays or X-rays that escape the decay site and will not form part of the cluster. Only 82\% of \pbten\ decays fall in this energy range. Of those, about 7.4\% are not selected because their measured cluster energy is $<$30\,keV due to pixel saturation. The corresponding pair selection efficiency is found to be $\epsilon_{\rm{pair}}=71.0\%$. Also, an average $\epsilon_{\rm{time}}=$65.1\% is obtained for this decay sequence.

The overall efficiency for detection of \sitwo\,--\ptwo\ (\pbten\,--\biten) decay sequences in our data set, $\epsilon_{\rm{X}}$=$\epsilon_{\rm{pair}}\epsilon_{\rm{time}}$, where X = Si and Pb, is determined to be $\epsilon_{\rm{Si}}$$=$49.2\% ($\epsilon_{\rm{Pb}}$$=$46.2\%).

\subsection{Limits on \sitwo\ and \pbten\  contamination}
\label{sec:results}

\begin{figure}[t!]
\begin{center}
\subfigure[\sitwo\,--\ptwo ]{
\includegraphics[width=0.475\textwidth]{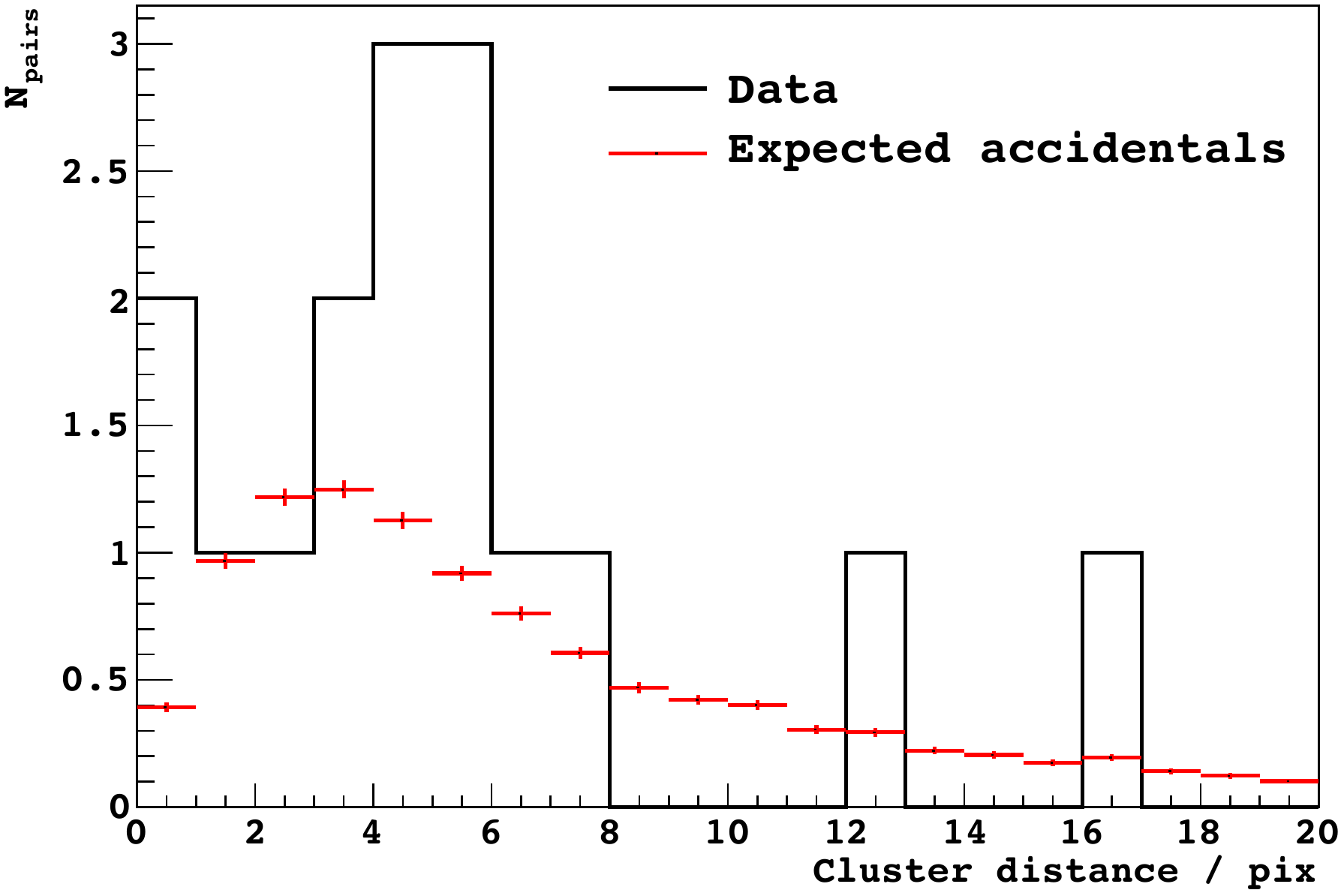}
\label{fig:distance_data}
}
\subfigure[\sitwo\,--\ptwo]{
\includegraphics[width=0.475\textwidth]{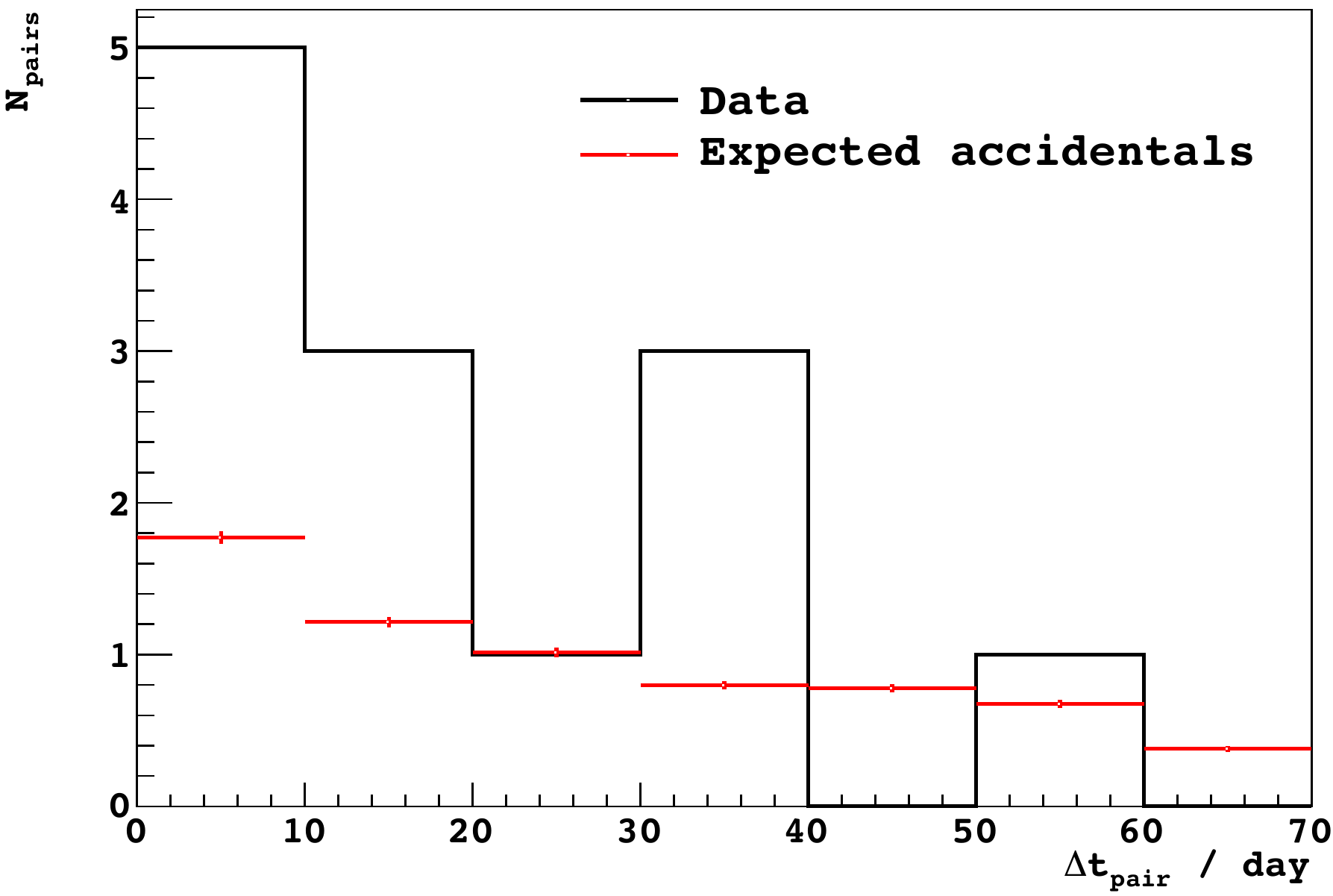}
\label{fig:time_data}
}
\end{center}
\caption{a)~Cluster distance distribution for the 16 \sitwo\,--\ptwo\ candidate pairs found in 56.8 days of data. b)~Distribution of the time separation $\Delta t_{\rm{pair}}$ of the 13 pairs with cluster distance $<7$ pixels. The red line is the expectation for accidental pairs.
}
\end{figure}

\begin{figure}[t!]
\begin{center}
\subfigure[\sitwo\,--\ptwo]{
\includegraphics[width=0.475\textwidth]{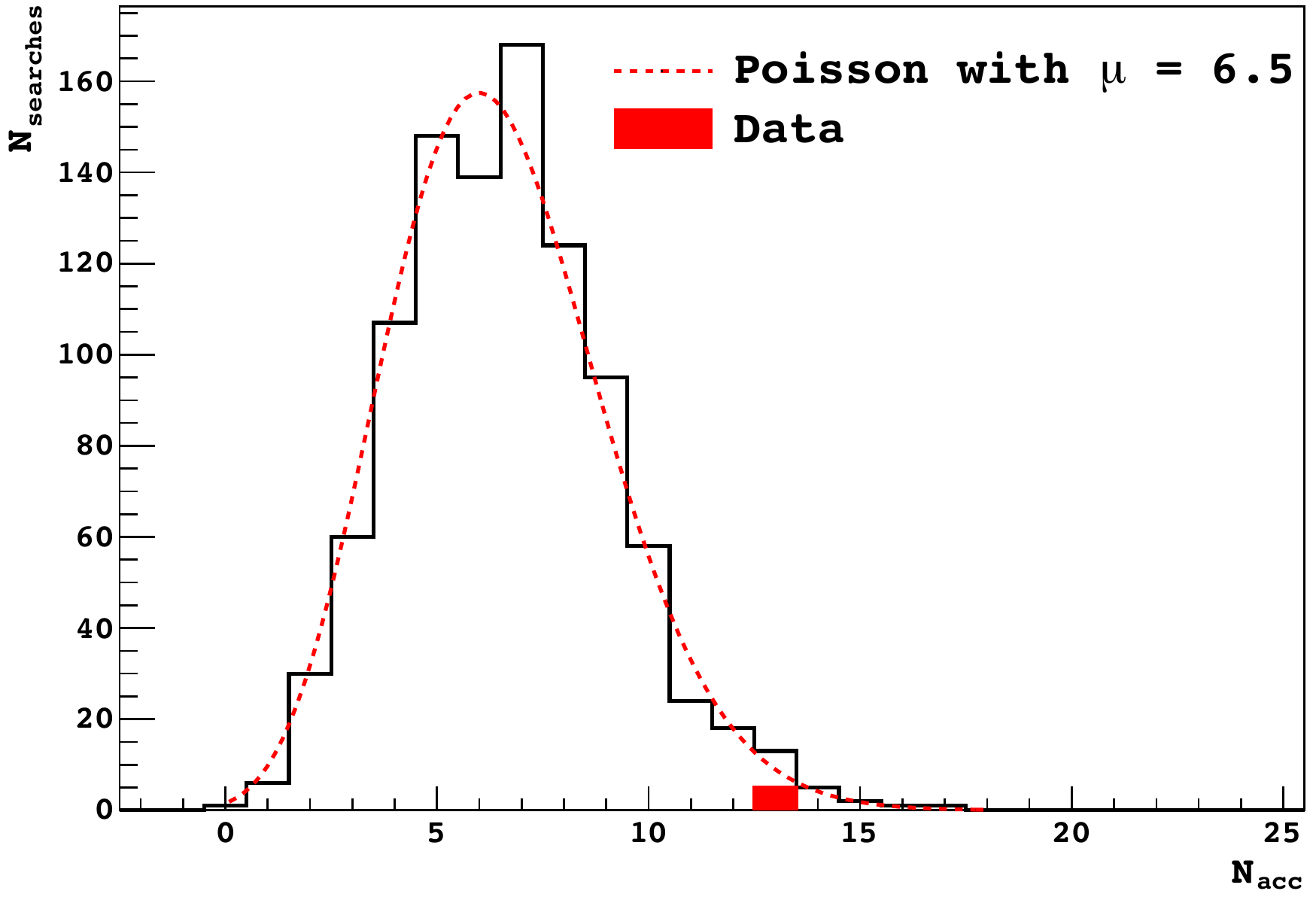}
\label{fig:si32_n1dist}
}
\subfigure[\pbten\,--\biten]{
\includegraphics[width=0.48\textwidth]{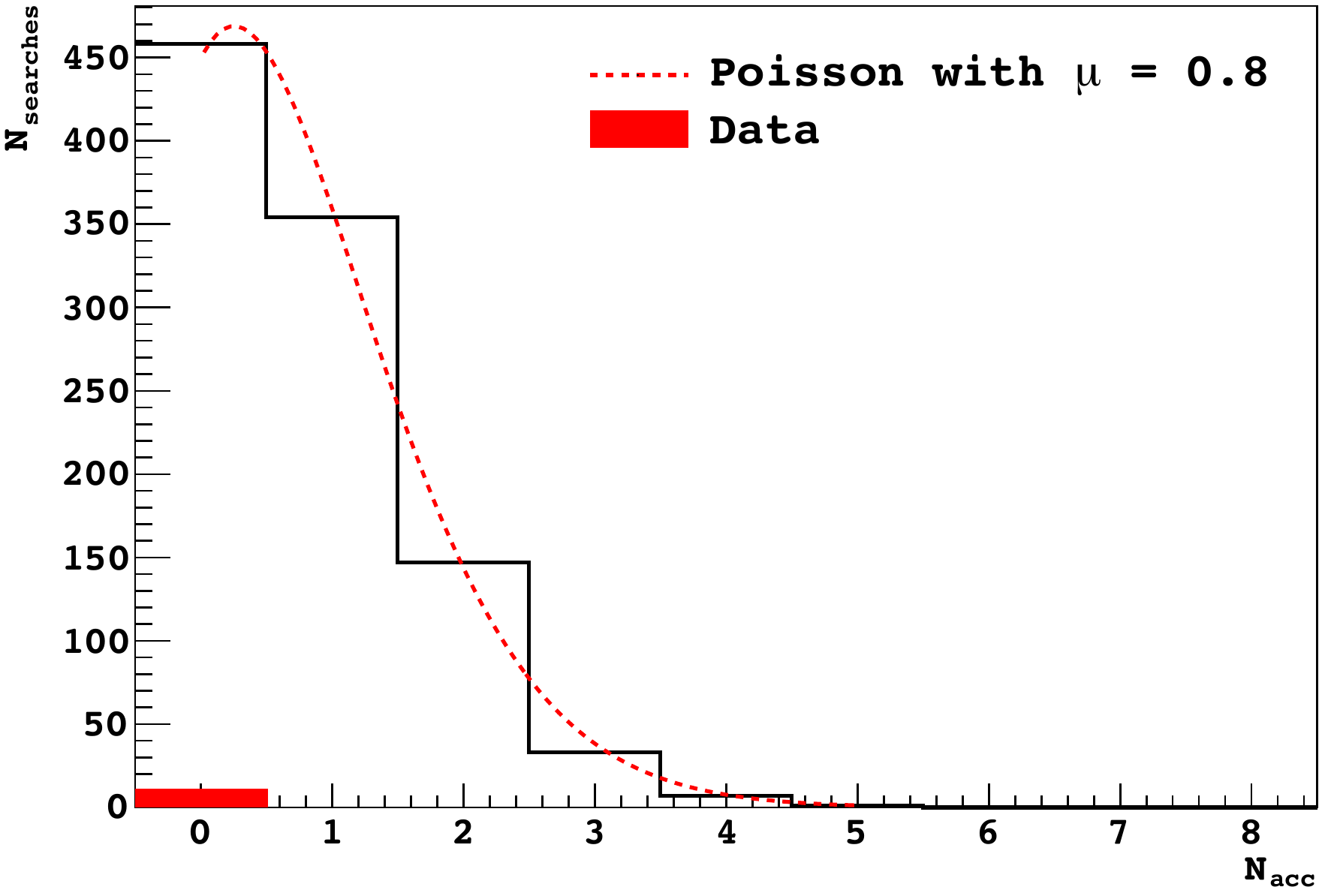}
\label{fig:pb210_n1dist}
}
\end{center}
\caption{Distribution of the number of accidental pairs with cluster distance $<$7 pixels for the  \sitwo\,--\ptwo\ (a) and \pbten\,--\biten\ (b) decay sequences. The distributions were obtained by performing 1000 searches over spatially randomized data images. The red-dashed line is the Poisson distribution corresponding to the mean of the histogram. The red entry is the number of candidate pairs found in data.
\label{fig:n1dist}}
\end{figure}

The search procedure described in Section~\ref{sec:procedure} was applied to the data. No candidate was found in the \pbten\,--\biten\ search. 
The \sitwo\,--\ptwo\ decay sequence criteria yielded 16 candidate pairs, 13 of which have a cluster distance $<$7 pixels (Figure~\ref{fig:distance_data}).
The  $\Delta t_{\rm{pair}}$ distribution of these 13 pairs is shown in Figure~\ref{fig:time_data}.  
Also shown are the corresponding distributions expected for purely accidental pairs, which were estimated directly from the data. For this purpose, we randomized the position of the clusters in all data images, effectively eliminating any spatial correlation between $\beta$s from decay sequences potentially present in the data. In the randomization process, the overall $x$-$y$ distribution of the clusters was maintained to avoid introducing a bias. Accidental pairs were then found in the set of randomized images by applying the search criteria. The procedure was repeated one thousand times, obtaining  for each randomized search the accidental pairs' distribution of cluster distance and $\Delta t_{\rm{pair}}$. The average of the one thousand distributions (red line) is shown in Figs.~\ref{fig:distance_data}-\ref{fig:time_data}. Within the limited statistics, an excess of pairs at small cluster distances is observed in data. Their $\Delta t_{\rm{pair}}$ distribution is compatible with the 14-day half-life of \ptwo.

To evaluate the significance of the excess in the  \sitwo\,--\ptwo\ search and establish a limit on \pbten\ contamination,  the probability distribution of the number of accidental pairs, $P_{\rm{acc}}$, was determined with the same spatial randomization procedure described above. For each randomized search, the number of accidental pairs with cluster distance $<$7 pixels, $N_{\rm{acc}}$, was obtained. A $N_{\rm{acc}}$ distribution was derived from one thousand randomized searches, and taken as an estimate of $P_{\rm{acc}}$. We verified with Monte Carlo simulations that the true $P_{\rm{acc}}$ is indeed recovered by this procedure. 
Figure~\ref{fig:n1dist} shows the $N_{\rm{acc}}$ distributions obtained for the  \sitwo\,--\ptwo\ and \pbten\,--\biten\ decay sequence searches.  These distributions are well represented by Poisson distributions with mean of 6.5 for the \sitwo\,--\ptwo\  search, and mean of 0.8 for  the \pbten\,--\biten\ search.  With this assumption, the 13 observed candidate pairs correspond to 1.2$<$$N_{\rm{Si}}$$<$15.3 (95\% CI)~\cite{Feldman:1997qc}, where $N_{\rm{Si}}$ is the number of estimated \sitwo\,--\ptwo\ decay sequences in the data. The null result for the \pbten\,--\biten\ search  corresponds to $N_{\rm{Pb}}$$<$2.5 (95\% CL)~\cite{Feldman:1997qc}, where $N_{\rm{Pb}}$ is the number of estimated decay sequences in the data.
 
The \sitwo\ and \pbten\ decay rates are then obtained as $N_{\rm{X}}/\epsilon_{\rm{X}}/\rm{T}/\rm{M_{CCD}}$, where $\epsilon_{\rm{X}}$ is given in Section~\ref{sec:efficiency}, T$=$56.8 d is the data live-time and $\rm{M_{CCD}}$$=$$2.9 \times 10^{-3}$~kg. We estimate a decay rate of $80^{+110}_{-65}$\,kg$^{-1}$\,d$^{-1}$ (95\% CI) for \sitwo\ in the CCD bulk. This result also establishes the detection of \sitwo\,--\ptwo\ pairs in our data set at 98\% CL. 
The derived upper limit on the \pbten\ decay rate in the CCD bulk is $<$33\,kg$^{-1}$\,d$^{-1}$ (95\% CL).

\begin{table}[t!]
\caption{Summary of results presented in this paper. All values are 95\% CL upper limits or intervals, except for the \poten\ surface rate, where the uncertainties are 1-$\sigma$. The two measurements of the \poten\ surface rate correspond to the two (back, front) CCD surfaces. For \ura\ and \tho, we quote the corresponding ppt contamination in parentheses.}
\centering
\begin{tabular}{|c|c|c|c|c|}\hline
Analysis &Isotope(s)	& Tracer  	& \underline{Bulk rate} 	& \underline{Surface rate}   \\
method &			&	for		&		kg$^{-1}$\,d$^{-1}$	&		cm$^{-2}$\,d$^{-1}$		\\  \hline 
$\alpha$& \poten		&	\pbten	&			$<$37				&	 0.011$\pm$0.004, 0.078$\pm$0.010			\\
spectroscopy & \urafour\ $+$ $^{230}$Th $+$ \radium\		&	\ura		&	$<$5 (4\,ppt)  &				--				\\
& $^{224}$Ra\,--\thoron\,--$^{216}$Po &	\tho		&		$<$15 (43\,ppt)  &		--	\\	\hline
$\beta$ spatial & \sitwo\,--\ptwo\		&	\sitwo		&	$80^{+110}_{-65}$	&			--			\\
coincidence &	\pbten\,--\biten\		&	\pbten		&			$<$33				&	--		\\ \hline
\end{tabular}
\label{tab:results}
\end{table}

\section{Conclusions and Outlook}
\label{conclusions}

We have presented novel analysis methods to measure radioactive contamination in the high-resistivity silicon CCDs of the DAMIC experiment. 
We exploited the unique signatures of $\alpha$ particles in CCDs to perform $\alpha$ spectroscopy and search for nuclides of the uranium and thorium chains. Also, we searched for \sitwo\ and \pbten\ in the bulk silicon of the CCD by identifying pairs of spatially correlated $\beta$ tracks compatible with \sitwo\,--\ptwo\  and \pbten\,--\biten\ decay sequences.

The results are summarized in Table~\ref{tab:results}. We placed stringent 95\% CL upper limits on the presence of radioactive contaminants in the silicon bulk.  The \ura\ and \tho\ decay rates were found to be $<$5\,kg$^{-1}$\,d$^{-1}$ and $<$15\,kg$^{-1}$\,d$^{-1}$, respectively. 
Also, we established an upper limit of $\sim$35\,kg$^{-1}$\,d$^{-1}$ (95\% CL) on the \pbten\ decay rate, obtained independently by $\alpha$ spectroscopy and the $\beta$ decay sequence search. In addition, we have measured the decay rate of \sitwo\ in the silicon bulk to be $80^{+110}_{-65}$\,kg$^{-1}$\,d$^{-1}$ (95\% CI). 
Since we detect single nuclear decays with high efficiency, our analysis methods have {near-}optimal sensitivity to radioactive contamination for a given exposure. In particular, the capability to identify \sitwo\,--\ptwo\  and \pbten\,--\biten\ decay sequences,  unique among particle detectors, allows us to measure these contaminations at levels that are orders of magnitude lower than would be possible with any other screening technique.

The rate of $\alpha$s on the surface of the CCDs was also measured, and results are reported in Table~\ref{tab:results}. Since the observed rates may be partly caused by contamination on surfaces surrounding the CCDs, these results should be taken as upper limits on the surface contamination of the CCDs.

These levels of radioactive contamination will allow DAMIC100 to probe WIMP-nucleon spin-independent interaction cross-sections as small as $10^{-5}$\,pb for WIMPs with masses as low as 2\,GeV/$c^2$.
In addition, material screening and handling procedures implemented for DAMIC100 should result in a tenfold reduction of the present background level, allowing for even more stringent limits on the radioactive contamination. Confirming the measurement, currently statistically limited, of \sitwo\ in high-resistivity silicon will be particularly relevant. In fact, the presence of the low energy $\beta$ decay of this cosmogenic isotope may impose additional constraints on the next generation WIMP searches with high-purity silicon detectors, including the identification of a source of silicon with low \sitwo\ content to fabricate the detector and the requirement of \sitwo\,--\ptwo\ single-event identification for background suppression.

\acknowledgments
The DAMIC Collaboration would like to thank SNOLAB and its staff for providing underground laboratory space and outstanding technical support, and Vale S.A. for hosting SNOLAB. We thank G. E. Derylo and K. R. Kuk for their contributions to the design, construction and installation of the detector. We are grateful to the following agencies and organizations for financial support: Kavli Institute for Cosmological Physics at the University of Chicago through grant NSF PHY-1125897 and an endowment from the Kavli Foundation, the Natural Sciences and Engineering Research Council of Canada, the Ontario Ministry of Research and Innovation, the Northern Ontario Heritage Fund, the Canada Foundation for Innovation, DGAPA-UNAM through grants PAPIIT No. IN112213 and No. IB100413, Consejo Nacional de Ciencia y Tecnolog\'{\i}a (CONACYT), M\'{e}xico, through grant No. 240666, the Swiss National Science Foundation through grant 153654, and the Brazilian agencies Coordena\c{c}\~{a}o de Aperfei\c{c}oamento de Pessoal de N\'{\i}vel Superior (CAPES), Conselho Nacional de Desenvolvimento Cient\'{\i}fico e Tecnol\'{o}gico (CNPq) and Funda\c{c}\~{a}o de Amparo \`{a}  Pesquisa do Estado de Rio de Janeiro (FAPERJ).
 
\bibliographystyle{jhep}
\bibliography{myrefs}

\end{document}